\documentclass[12pt]{article}

\usepackage{times,color,bm}
\usepackage{epsfig,amsmath,amssymb,amsfonts,mathrsfs,braket}
\usepackage{subfigure,graphicx,epstopdf}
\usepackage{tikz,cite}
\usetikzlibrary{matrix,fit}
\usepackage[colorlinks]{hyperref}
\hypersetup{linkcolor = {blue},	citecolor = {blue},	urlcolor = {blue}}

\newenvironment{sciabstract}{%
	\begin{quote} \bf}
	{\end{quote}}

\newcounter{lastnote}

\title{Protecting Quantum Superposition and Entanglement with Photonic Higher-Order Topological Crystalline Insulator}

\author
{Yao Wang,$^{1,2,\dagger}$ Bi-Ye Xie,$^{3,4,\dagger}$ Yong-Heng Lu,$^{1,2}$ Yi-Jun Chang,$^{1,2}$\\
	Hong-Fei Wang,$^{3,4}$ Jun Gao,$^{1,2}$ Zhi-Qiang Jiao,$^{1,2}$ Zhen Feng,$^{1,2}$ Xiao-Yun Xu,$^{1,2}$\\
	Feng Mei,$^{5,6}$ Suotang Jia,$^{5,6}$ Ming-Hui Lu,$^{3,4,7,8}$ and Xian-Min Jin$^{1,2}$\\
	\\
	\normalsize{$^1$Center for Integrated Quantum Information Technologies (IQIT), School of Physics}\\
	\normalsize{and Astronomy and State Key Laboratory of Advanced Optical Communication Systems}\\
	\normalsize{and Networks, Shanghai Jiao Tong University, Shanghai 200240, China}\\
	\normalsize{$^2$CAS Center for Excellence and Synergetic Innovation Center in Quantum Information}\\
	\normalsize{and Quantum Physics, University of Science and Technology of China, Hefei, Anhui 230026, China}\\
	\normalsize{$^3$National Laboratory of Solid State Microstructures, Nanjing University, Nanjing 210093, China}\\
	\normalsize{$^4$Department of Materials Science and Engineering, Nanjing University, Nanjing 210093, China}\\
	\normalsize{$^5$State Key Laboratory of Quantum Optics and Quantum Optics Devices,}\\
	\normalsize{Institute of Laser Spectroscopy, Shanxi University, Taiyuan, Shanxi 030006, China}\\
	\normalsize{$^6$Collaborative Innovation Center of Extreme Optics,}\\
	\normalsize{Shanxi University, Taiyuan, Shanxi 030006, China}\\
	\normalsize{$^7$Jiangsu Key Laboratory of Artificial Functional Materials, Nanjing 210093, China}\\
	\normalsize{$^8$Collaborative Innovation Center of Advanced Microstructures,}\\
	\normalsize{Nanjing University, Nanjing 210093, China}\\
	\normalsize{$^\dagger$These authors contributed equally to this work}\\
}

\date{}

\begin{document}
	\baselineskip24pt
	
	\maketitle
	
	\begin{sciabstract}
Higher-order topological insulator, as a newly found non-trivial material and structure, possesses a topological phase beyond the bulk-boundary correspondence. Here, we present an experimental observation of photonic higher-order topological crystalline insulator and its topological protection to quantum superposition and entanglement in a two-dimensional lattice. By freely writing the insulator structure with femtosecond laser and directly measuring evolution dynamics with single-photon imaging techniques, we are able to observe the distinct features of the topological corner states in $C_4$ and $C_2$ photonic lattice symmetry. Especially, we propose and experimentally identify the topological corner states by exciting the photonic lattice with single-photon superposition state, and we examine the protection impact of topology on quantum entanglement for entangled photon states. The single-photon dynamics and the protected entanglement reveal an intrinsic topological protection mechanism isolating multi-partite quantum states from diffusion-induced decoherence. The higher-order topological crystalline insulator, built-in superposition state generation, heralded single-photon imaging and quantum entanglement demonstrated here link topology, material, and quantum physics, opening the door to wide investigations of higher-order topology and applications of topological enhancement in genuine quantum regime.\\
	\end{sciabstract}
	
	\subsection*{Introduction}
	Topological phase, possessing intriguing bulk and edge properties, plays a fundamental role in understanding matter~\cite{Hall1,review1,review2,review3,review4} and constructing artificial devices. It displays an extraordinary robustness to smooth changes in material parameters or disorders and endows the system with inherent protection~\cite{review1, review2}. In addition to fundamental physics, the extraordinary robustness of topological phase has found promising applications for inherently fault-tolerant quantum simulation and quantum computing~\cite{computing}. In the past decades, topological robust phases have been widely investigated in various systems including condensed-matter~\cite{review1, review2}, ultracold matter~\cite{Cold1, Cold2, Cold3}, phononic~\cite{Topo_weyl_1,XBY4,XBY1} and photonic systems~\cite{review3, review4, photon1, photon2, photon3}.
	
	Compared with other information carriers, photon is easier to be created and detected, and the topological photonics is rapidly developed in recent years. Together with the three-dimensional femtosecond laser direct writing technique~\cite{fabri_1}, we are able to construct photonic waveguide lattice on a photonic chip in a physically scalable and structurally designable fashions~\cite{LargestChip, 3Dscalable, PIT_Gap}. The constructed photonic waveguide array provides a convenient platform for investigating the topological phase, especially, for the higher-order topological insulators (HOTIs) and topological crystalline insulators (TCIs) in a high complexity~\cite{TCI}.
	
	HOTIs have been recently proposed as a novel topological phase of matter with unconventional bulk-boundary correspondence~\cite{HOTI1, HOTI2, HOTI3, HOTI4, HOTI5, HOTI7, HOTI8, HOTI9, HOTI10, HOTI12, HOTI13, HOTI14, HOTI15, corner, XBY3, QP1, quadrupole, C3}. An $N$th-order topological insulator has topological boundary state with codimension $N$~\cite{CO1,CO2}. Generally, there are two ways to achieve HOTIs. The first one is to use topological multipole insulators which were firstly proposed by W. Benalcazar et al.~\cite{HOTI1,QP1} and lately experimentally realized in mechanics~\cite{HOTI2}, microwaves~\cite{HOTI3}, electric circuits~\cite{HOTI4}, phononics~\cite{HOTI15} and coupled-resonator optical waveguides quadrupole~\cite{quadrupole}. The other approach is to utilize topological crystalline insulators with quantized bulk polarization which was theoretically proposed by considering tight-binding models~\cite{HOTI8,CO1} and experimentally achieved in photonics~\cite{HOTI9, HOTI10, XBY3, C3} and phononics~\cite{HOTI12, HOTI13, HOTI14}. However, the investigations of the intriguing lower dimensional topological boundary states in HOTIs down to single-particle level as well as their quantum characters have not yet been explored. The impact of topology on quantum entanglement for entangled photon states also attracts the interest, which inspire a fascinating and elegant combination of topological photonics and quantum photonics. 
	
	Here, we theoretically and experimentally investigate the higher-order topological phases and the topological protection to quantum superposition and entanglement in two-dimensional photonic lattice via single-photon dynamics. In contrast to all-dielectric photonic crystals, the zero-dimensional corner state generated in the second-order topological phases is more sensitive to the crystalline symmetry of photonic lattice. We experimentally demonstrate two ways of identifying the crystal-symmetry-dependent topological corner states in the photonic lattice, with single-site and superposition-state injection. The topological corner state is found being able to preserve single-photon superposition state and the quantum entanglement against diffusion-induced decoherence, representing an intrinsic topological protection mechanism for multi-partite quantum states.\\
	
	\subsection*{Experimental implementation and results}
\begin{figure}[htbp]
		\centering
		\includegraphics[width=1.0\linewidth]{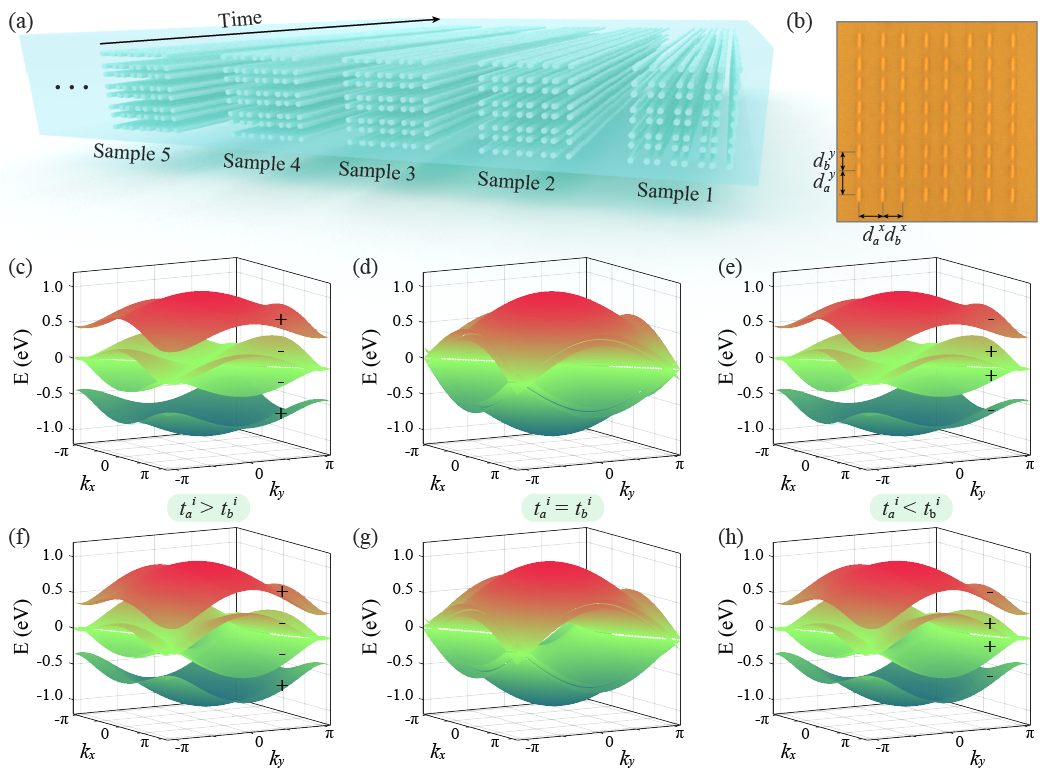}
		\caption{\textbf{Schematic of topological photonic chip and the band structure of the lattice.} \textbf{(a)} The topological lattices are integrated in a single photonic chip. \textbf{(b)} The microgram of the photonic lattice crosssection. \textbf{(c-h)} The band structures of $C_2$ and $C_4$ symmetric lattices. There is a band inversion process between two gapped phases separated by gapless configurations for both $C_2$ (c-e) and $C_4$ (f-h) symmetric lattices when $t^i_b=t^i_a$.}
		\label{f1}
	\end{figure}

	\paragraph*{Integrated topological lattice.} In our experiment, we fabricate the higher-order topological insulators in alkaline earth boro-aluminosilicate glass using femtosecond laser direct writing technique (see Methods)~\cite{fabri_1}. We integrate various samples in a photonic chip, as shown in Fig.~\ref{f1}(a). The constructed two-dimensional lattice contains $8\times8$ sites and the evolution distances vary from 10 to 30 mm with step of 5 mm. Our photonic lattice mimics the two dimensional (2D) extended Su-Schrieffer-Heeger (SSH) model~\cite{SSH1}, the corresponding Hamiltonian in the momentum space can be expressed as
	\begin{equation}         
		\mathcal{H}(\bm{k})=
		\left(                 
		\begin{array}{cccc}   
			0 & h_{12} & h_{13} & 0\\  
			h_{12}^\ast & 0 & 0 & h_{24}\\ 
			h_{13}^\ast & 0 & 0 & h_{34}\\ 
			0 & h_{24}^\ast & h_{34}^\ast & 0 \\
		\end{array}
		\right),
		\label{eq:1}
	\end{equation}
	where $h_{12}=t^x_a+t^x_b\mathrm{exp}(\mathrm{i}k_x)$, $h_{13}=t^y_a+t^y_b\mathrm{exp}(\mathrm{-i}k_y)$, $h_{24}=t^y_a+t^y_b\mathrm{exp}(-\mathrm{i}k_y)$, $h_{34}=t^x_a+t^x_b\mathrm{exp}(\mathrm{i}k_x)$, and ${\bm k}=(k_x, k_y)$. The parameter $t^i_a$ ($t^i_b$) with $i=x,y$ represents the intra-cell (inter-cell) coupling strength along $i$-direction. The coupling strength is modulated by the separation between two sites, as shown in Fig.~\ref{f1}(b), the corresponding separation distance for $t^i_a$ ($t^i_b$) is $d^i_a$ ($d^i_b$). For our waveguide array, the coupling between nearest-neighbour sites are positive which means $t^i_j>0$ for $i=x,y$ and $j=a,b$ and the coupling between next-nearest-neighbour (or higher-order-neighbour) sites are exponentially suppressed~\cite{2ndC} which ensures the validity of the tight-binding approximation.
	
	By diagonalizing the Hamiltonian in Eq.(\ref{eq:1}), we can obtain the band structure of the photonic lattice [see Figs.\ref{f1}(c)-\ref{f1}(h)]. The competition between $t^i_a$ and $t^i_b$ with $i=x, y$ determines the existence of the band gap where the band gap closes at the topological phase transition point $t^i_a=t^i_b$. On the other hand, if $t^x_j \neq t^y_j$ ($t^x_j = t^y_j$) for $i=a,b$, the system described by the above Hamiltonian always have $C_2$ ($C_4$) rotation symmetry. The band structures of $C_4$ and $C_2$ symmetric lattices with different coupling parameters are presented in Figs.\ref{f1}(c)-\ref{f1}(e) and Figs.\ref{f1}(f)-\ref{f1}(h) respectively, where the corresponding relationship are $t^i_a>t^i_b$, $t^i_a=t^i_b$, and $t^i_a<t^i_b$. A band inversion process has been observed in both cases, indicating a topological phase transition when $t^i_a=t^i_b$.

	\paragraph*{Topological corner index and filling anomaly.} The topological photonic lattices with $C_4$ and $C_2$ crystalline symmetry both are in the second-order topological phases with 1D topological edge states and 0D topological corner states (see Methods). To investigate the corner physics of the second-order topological insulator (SOTI), we explore the rotation symmetry group representations at high symmetry points in Brillouin zone~\cite{TCI1,HOTIC}. Due to the symmetries, the eigenstates of the Hamiltonian can be chosen as the common eigenstates of the rotation operators $\hat{R}_n$ ($n=2,4$) with corresponding eigenvalues $\Pi^{(n)}_p=e^{2\pi i(p-1)/n}$ with $p=1,2,...,n$. By comparing the rotation eigenvalues at high symmetry points to those of atomic insulators, we can determine whether the system is topologically non-trivial. Specifically, we can define the topological index as
	\begin{equation}
		[\Pi^{(n)}_p]=\#\Pi^{(n)}_p-\#\Gamma^{(n)}_p,
		\label{index}
	\end{equation}
	where $\#\Pi^{(n)}_p$ is the number of bands below the band gap with rotation eigenvalues $\Pi^{(n)}_p$. $\Pi^{(n)}$ stand for high symmetric point $X$, $M$ and $\Gamma$ in $C_n$ symmetric systems. When $[\Pi^{(n)}_p]$ is non-zero, the system is a topological insulator (denoted as the obstructed atomic insulator).
	
	Theoretically the indices defined in Eq.\ref{index} can fully characterize the higher-order topological properties of the systems. However, considering the time-reversal symmetry and the fact that the number of bands below the gap is constant across the Brillouin zone, these indices are not independent to each other, thus we can drop the redundant indices. For $C_4$ symmetric lattices, the indices are $[X_1]$, $[M^{(4)}_1]$ and $[M^{(4)}_2]$ while for $C_2$ symmetric lattices, the indices are $[X_1]$, $[Y_1]$ and $[M^{(1)}_1]$. This approach is based on the theory of topological crystalline insulators~\cite{HOTIC} and similar to the recently proposed topological symmetry indicators~\cite{SI1,SI2,SI3}.

	We further determine the values of these topological indices by investigating the coupling configurations of lattice. We find that it is $\frac{1}{4}$ fractionalized corner states at each of four corners for our $C_4$ symmetric lattice, but $\frac{1}{2}$ fractionalized corner states at each of four corners for $C_2$ symmetric lattice (see Methods for details). The topological corner index captures the corner physics more precisely and directly than the bulk polarization. Moreover, it provides the fractionalization of the photonic eigenstates. Due to the Abelian additive structure of TCI, we can theoretically construct TCIs with other topological corner index by using these primitive generators and even for the fragile TCIs that do not admit Wannier representations \cite{HOTIC}.
	
	\paragraph*{Chiral symmetry and the separation between corner states and bulk states.} Different from the all-dielectric photonic crystals~\cite{XBY3}, the energy levels between corner states and bulk states of our system is able to manifest the difference between $C_2$ and $C_4$ rotation symmetry. In 1D SSH model~\cite{SSH_1,SSH_2,SSH_3}, due to the staggered coupling strengths between inter-cell and intra-cell sites, there is a sublattice symmetry which is also the chiral symmetry of the Hamiltonian since it satisfies $\Gamma H(\bm{k})\Gamma^{-1}=-H(\bm{k})$, where $\Gamma$ presents operation of chiral symmetry. It restricts the band structures (or more precisely, the eigenvalues) to be symmetric with respect to ``zero energy". In 2D cases, due to the exponentially suppressing of the higher-order couplings, the chiral symmetry is preserved in $C_4$ symmetric lattice while being broken in $C_2$ symmetric lattice. Therefore, in $C_4$ symmetric case, the corner states are pinned on the zero-energy level and embedded into the bulk states. In fact, these corner states are actually the higher-order version of bound states in the continuum, the details analysis can be found in Supplementary Materials. On contrary, in $C_2$ symmetric case, the corner states are not restricted to the zero-energy level and can be separated from the bulk states and edge states. The above analysis can be confirmed by our following experimental observation.
	
\begin{figure}[htbp]
		\centering
		\includegraphics[width=0.9\linewidth]{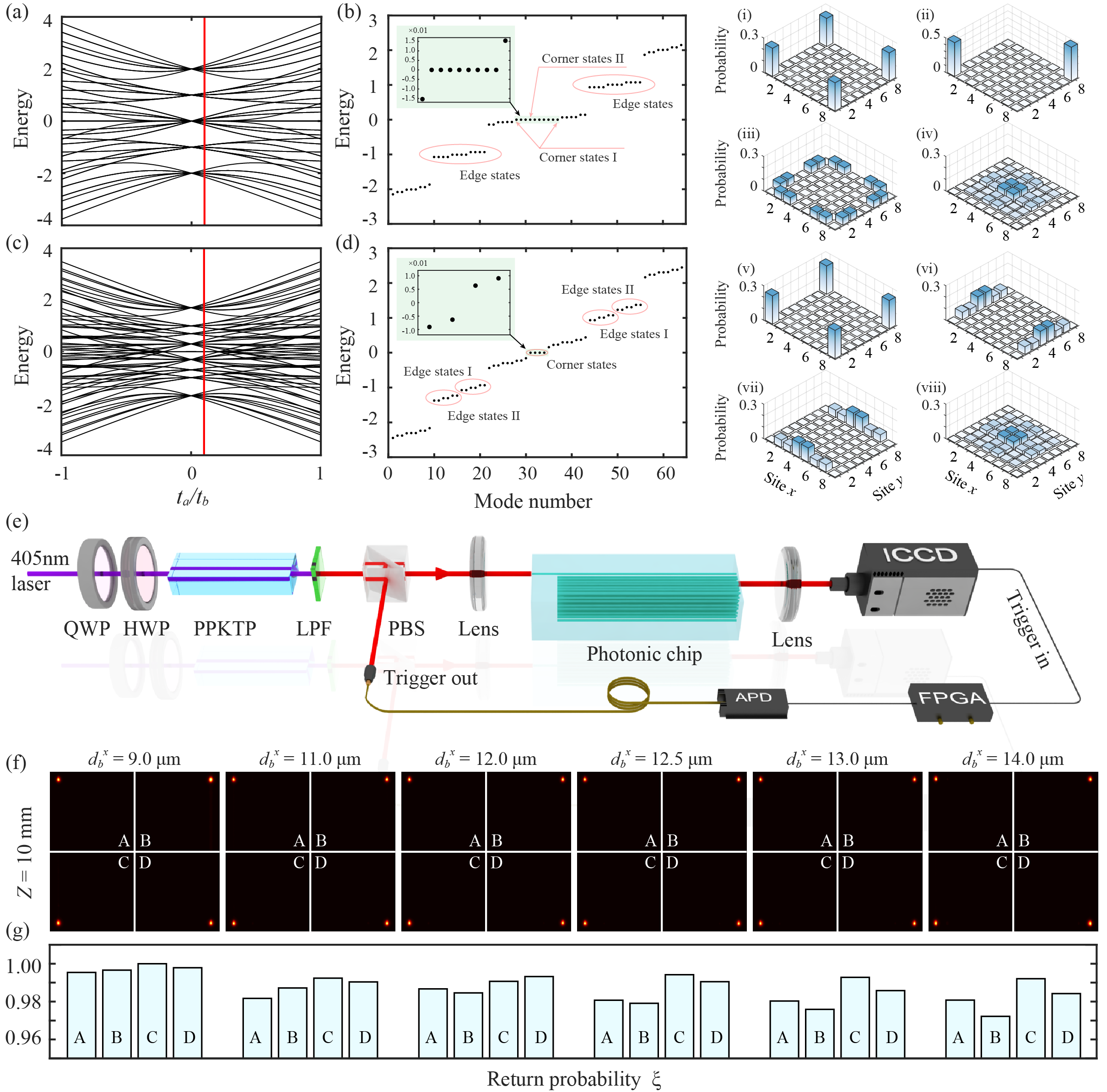}
		\caption{\textbf{The spectrum of finite lattice and measured single-photon distribution probability of corner states.} \textbf{(a-d)} The spectrum of the finite $C_2$ and $C_4$ symmetric lattices. The corner states and edge states behave differently for the $C_2$ and $C_4$ symmetric lattices. In (b), The spatial distribution of corner state I/II is presented in inset (i)/(ii). The spatial distribution of edge state I/II in (d) is shown in inset (vi)/(vii). The red lines in (a) and (c) point out the parameter $t_a/t_b$ picked in (b) and (d). \textbf{(e)} Schematic of experimental setup. The heralded single photons generated from the PPKTP crystal are injected into the lattices after being focused and then collimated by a lens, and collected at the output facet by an ICCD, meanwhile, the heralding photon act as the trigger. HWP: half-wave plate, QWP: quarter wave plate, LPF: long-pass filter, APD: avalanche photodiode. \textbf{(f-g)} The measured single-photon distribution probability and return probability of the corner states. The results of the other cases with different evolution distance can be found in Supplementary Materials. The parameters of lattice are adopted as $d_a^y=22\ \mu m$, $d_b^y=9\ \mu m$, $d_a^x=22\ \mu m$, and the $d_b^x$ is picked as marked in the figures. The first sample is $C_4$ symmetric lattice and the others are $C_2$ symmetric lattices.}
		\label{f2}
	\end{figure}

	\paragraph*{Corner states in finite lattice.} For the finite lattice, the edge states and the corner states manifest due to the dimensional hierarchy of topological phases. We show the spectrum of the $C_4$ and $C_2$ symmetric lattices in Figs.\ref{f2}(a)-\ref{f2}(d) respectively. For the finite $C_4$ symmetric lattice, we show the spectrum as the function of $t_a/t_b$ varying from -1 to 1 in Fig.~\ref{f2}(a), and the detailed spectrum for the case of $t_a/t_b=0.1$ in Fig.~\ref{f2}(b). We further show the spatial distribution of the zero-energy corner modes and the edge modes in the insets (i)-(iv) in Fig.~\ref{f2}, which can be defined by
	\begin{equation}
		D_n(E)=\sum_{m}\delta(E-E_m)|\varphi_n^{(m)}|^2,
	\end{equation}
	where $E_m$ is the energy of the $m$th eigenstate $\varphi_n^{(m)}$. As the spatial distribution shown, the photon is confined at the four or two corners of lattice under the norm of the zero-energy corner states [see insets (i-ii)]. The edge states imply that the photon occupies the boundaries of the lattice with high probability [see insets (iii)]. In contrast, the photon only distributes in the bulk of the lattice for the bulk states [see insets (iv)]. 
	
	For the finite $C_2$ symmetric lattice, the spectrum behaves differently with the $C_4$ symmetric lattice. There are no zero-energy bulk states near the corner states, and no eigen-diagonal corner state among the four corner eigenstates as the $C_2$ symmetric lattice. Moreover, for the $C_2$ lattice, the corner states are two-fold degenerate as shown in insert of Fig.~\ref{f2}(b), which is characterized by $Q^{(2)}_c=\frac{1}{2}$. The edge states have the same symmetry as the lattice, as a result, there are two kind eigen edge states in $C_2$ symmetric lattice [see insets (vi-vii)].
	
	\paragraph*{Identifying the corner states.} As analyzed above, the corners state behaves as that the single-photon probability distribution mainly occupies on the corners of lattice. According to the principle of quantum superposition, if we inject the single photon into the lattice from one of the corners, only the corner states mainly maintain the evolution of single photon in lattice, though all the states are excited. In this case, the single photon will be confined in the excited corner of lattice, which is the result of the superposition of all the zero-energy corner states. Such evolution is stable due to that these corner states are robust to the structure disorder and are decoupled from the bulk bands.
	
	In experiment, the parameters are adopted as $d^x_a=d^y_a=22\ \mu$m, $d^x_b=d^y_b=9\ \mu$m for the $C_4$ symmetric lattice. We inject the single photons into each corner and capture the single-photon distribution probability after different evolution distance varying from 10 to 30 mm with a step of 5 mm [see Fig.~\ref{f2}(e) and Methods for details]. The measured single-photon distribution probability with evolution distance of 10 mm are shown in Fig.~\ref{f2}(f), where the single photon almost only occupies on the excited site. To quantify the localization of the outgoing single-photons distribution, we define the generalized return probability as $\xi=\sum_{k-w}^{k+w}I_i/\sum_{1}^{n}I_i$, where the $n$ is the site number of lattice. The quantity $\xi$ quantifies the probability of the single photon remaining within a small width $w$ from the injected site $k$. For the corner states, the photon is expected to be localized in only the corner site, then the accumulated width $w=0$. Such that the return probability of corner state is $\xi=\sum_{k}I_i/\sum_{1}^{n}I_i$. In our experiment, as shown in Fig.~\ref{f2}(g), the return probability of corner state $\xi$ maintains in high value close to 1 and doesn't change with the increase of evolution distance. We modulate the $d^x_b$ varying from 11 to 14 $\mu$m, the corresponding $t_a/t_b$ varies from 0.08 to 0.22, and keep other parameters unchanged. Five $C_2$ symmetric lattices are integrated in one photonic chip. The measured results are shown in Fig.~\ref{f2}(f)-\ref{f2}(g), which accord to the $C_4$ symmetric lattice as analyzed in theory.
	
	\paragraph*{Compatibility of corner state and single-photon superposition state.} As we known, the quantum superposition, as the most fundamental principle of quantum mechanics, allows the single photon be in more than one state simultaneously, which is called single-photon superposition state. For example, if there are four configurations labeled by $\ket{\psi_i}$, $i=1,2,3,4$, the most general single-photon superposition state would be

	\begin{equation}
		\ket{\psi}=\sum_{1}^{4}c_i\ket{\psi_i},
	\end{equation}
	where the coefficients are complex numbers describing the probability of each configuration. With the help of such a single-photon superposition state, we are able to determinately identify the corner states without exciting all states.
	
\begin{figure}[htbp]
		\centering
		\includegraphics[width=1.0\linewidth]{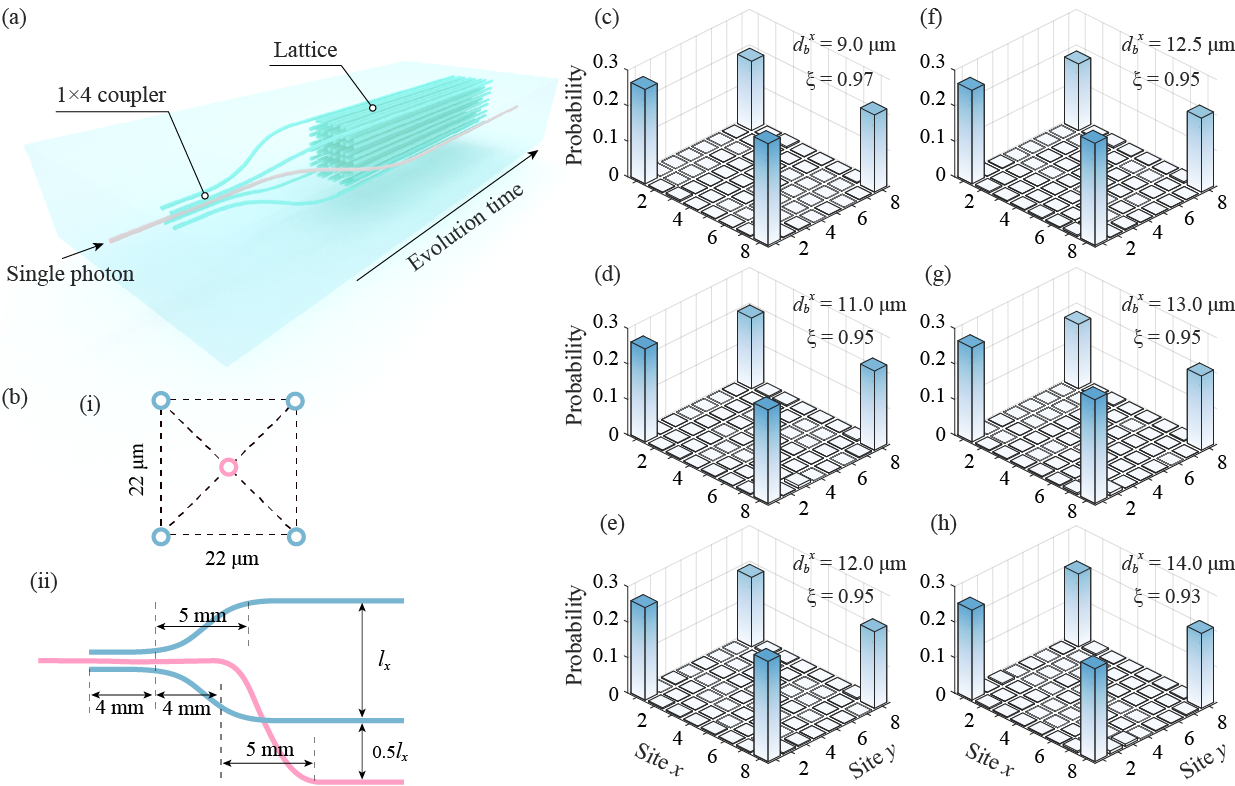}
		\caption{\textbf{The 1$\times$4 coupler and the measured single-photon distribution probability of corner states} \textbf{(a)} Schematic of the photonic lattices. A 1$\times$4 coupler is designed before the lattice. \textbf{(b)} Schematics of the details of the structures of the 1$\times$4 coupler of the cross section (i) and side section (ii). \textbf{(c-h)} The experimental results of corner states by exciting the lattices with single-photon superposition state. The parameters of lattice are adopted as $d_a^y=22\ \mu m$, $d_b^y=9\ \mu m$, $d_a^x=22\ \mu m$, and the $d_b^x$ is picked as marked in the figures. The first sample is $C_4$ symmetric lattice and the others are $C_2$ symmetric lattices. The results of the other cases with different evolution distance can be found in Supplementary Materials.}
		\label{f3}
	\end{figure}

	As shown in Fig.~\ref{f3}(a)-\ref{f3}(b), we obtain the single-photon superposition state as $\ket{\psi}=\frac{1}{2}\sum_{1}^{4}\ket{\psi_i}$ by injecting a single photon into a 3D 1$\times$4 photonic coupler [see Methods for details]. Subsequently, the prepared single photon is split into four corners of lattice to generate the desired superposition states, and the distribution probability of photon in lattice is identical to the zero-energy corner state. In the language of quantum mechanics, the system now is in the eigenstate and the single-photon superposition state in lattice will be maintained due to the orthogonality among eigenstates. With the same parameters, five kinds lattice are fabricated and integrated with the 3D 1$\times$4 photonic coupler on one chip. In Fig.~\ref{f3}(c)-\ref{f3}(h), we show the measured single-photon distribution probability and the return probability. The output probability distribution of single photon follows the distribution of corner states and is robust under the change of structure parameters of the systems. 
	
	 Meanwhile, the result also implies that the evolution of the single-photon superposition state is also protected by the topological corner mode against the structure disorder and diffusion induced photon loss. The successful observation of topological protected single-photon superposition state here adds one more key element into the toolbox of quantum topological photonics. Together with the previous work demonstrating the topologically protected quantum states and source~\cite{Qutop_1photon,Qutop_source,Qutop_interfere,Qutop_2photon_1,Qutop_2photon_2,Qutop_entanglement}, the robust features against photon loss induced by diffusion and the structure disorder promise that the topological protected single-photon superposition state is able to provide a valuable resource for quantum information processing.
	
\begin{figure}[htbp]
		\centering
		\includegraphics[width=1.0\linewidth]{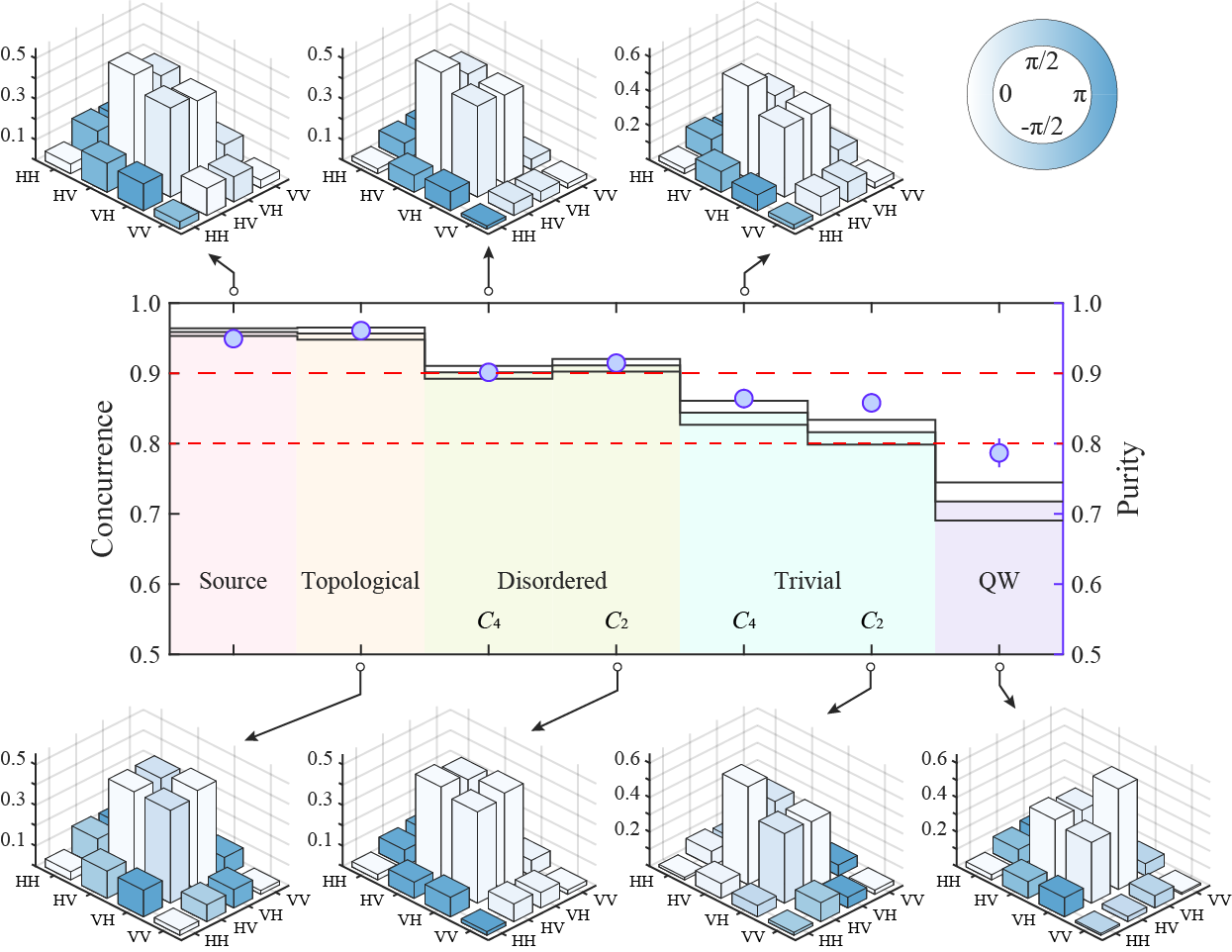}
		\caption{\textbf{Measured entanglement in higher order topological photonic chip.} Two-photon polarization entangled state is generated via the process of spontaneous parametric down-conversion. One of the two entangled photons is injected into the lattices, and the state tomography is conducted for different lattices. For the topological case and the disordered lattices, the measured values of concurrence and purity go well beyond 90\%. There is a obvious drop for the trivial case lower than 90\% and quantum walk (the gapless case) lower than 80\%. In the results of  reconstruct the density matrix, the modulus and argument of the matrix elements are represented by the height and color of the bars respectively.}
		\label{F_Entang}
	\end{figure}

	\paragraph*{Protecting the entanglement.} It is interesting to examine the impact of topological phase on quantum entanglement for entangled photon states~\cite{Qutop_2photon_2, Qutop_entanglement}. In our experiment, we inject the entangled photons into the corner state of the photonic lattice and reconstruct the density matrix of the output states. We calculate the concurrence and purity of the measured two-qubit entangled state to quantify the entanglement of the output photons, where the high concurrence and purity indicate the high entanglement quality and quantum state respectively. As shown in Fig.~\ref{F_Entang}, the entangled photons in the corner state preserve the high concurrence and purity beyond 90\%, even after introducing the disorder. In contrast, the entanglement in the trivial cases tends to corruption, which concurrence and purity drop to lower than 90\%. Especially, we fabricate a large-scale two-dimension uniform lattice with $21\times21$ sites to value the entanglement in the quantum walk. The concurrence and purity of the entangled state in the quantum walk drop to lower than 80\% after evolution distance of $z = 11$ mm, which lose the possibility of in practical applications.
	
\begin{figure}[htbp]
		\centering
		\includegraphics[width=1.0\linewidth]{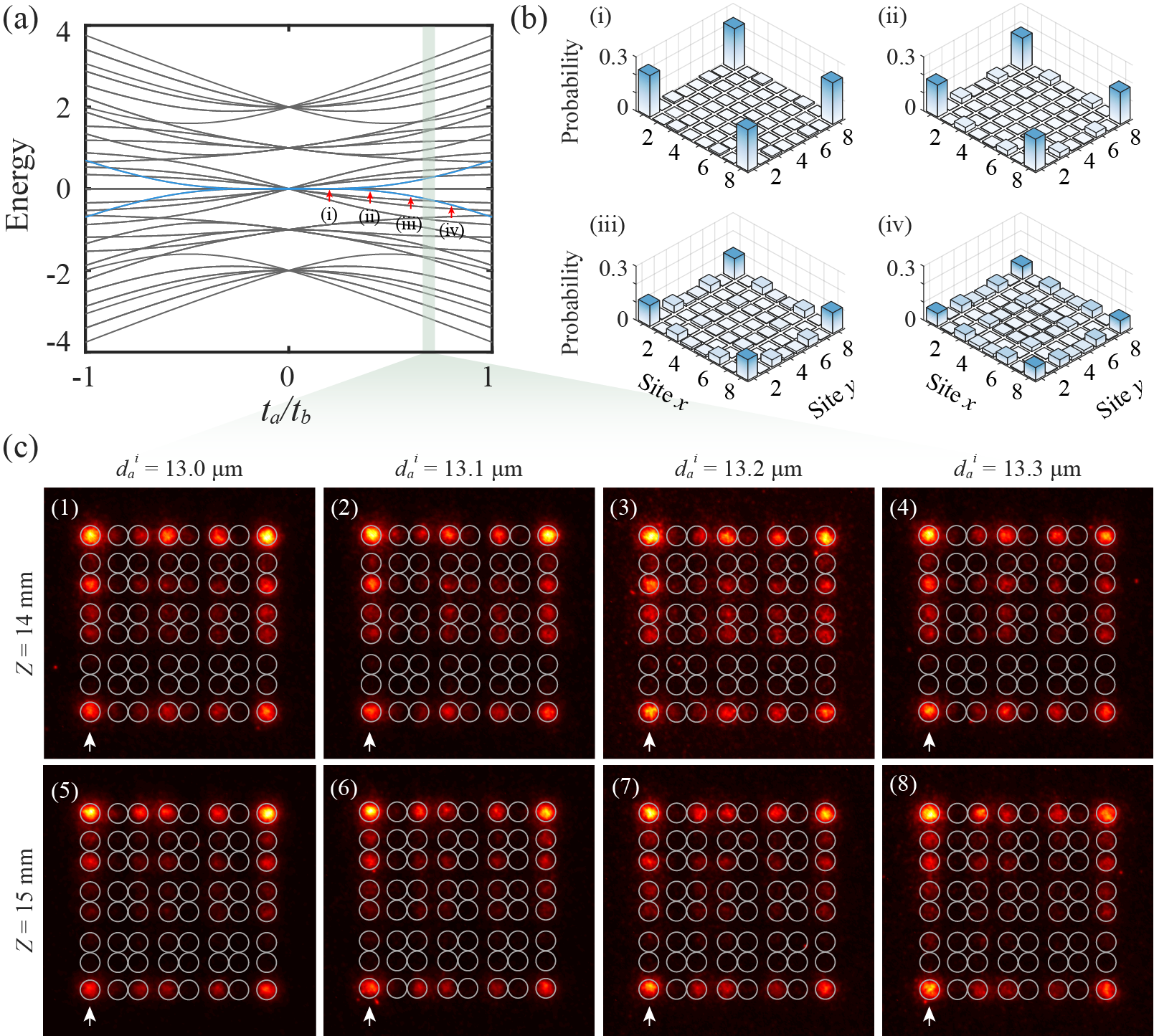}
		\caption{\textbf{Measured result of corer state with large value of $|t^i_a/t^i_b|$.} \textbf{(a)} The spectrum of the lattices as the function of $|t^i_a/t^i_b|$. The green region indicates the range of parameter adopted in experiment. \textbf{(b)} The field of corner states tends to distribute in the edge and bulk of the lattices with the increase of $|t^i_a/t^i_b|$. \textbf{(c)} The experiment results. The parameter $d_b^x=d_b^y$ is fixed as $11\ \mu$m and $d_a^x=d_a^y$ are set varying from 13 (insets 1 and 5) to $13.3\ \mu$m (insets 4 and 8) with step of $0.1\ \mu$m for devolution distance of 14 (insets 1-4) and 15 mm (insets 5-8). The white arrows point out the excited sites of the lattices.}
		\label{f4}
	\end{figure}

	\paragraph*{The finite gap effect.} As the spectrum shown in Fig.~\ref{f4}(a), the degenerate corner modes will transit to non-degenerate bulk states with the increase of $t_a/t_b$ for the topological phase. This is the consequence of the finite gap effect. Under the same size of lattice, when $t_a/t_b$ approaches to $1$, the gap size between the first band and second band (or the third band and the forth band) approaches to zero. Since the gap size is proportional to the mass term in the effective Hamiltonian which is exponentially proportional to the decay rate of the interface states away from the corners, the photon in corner states will tend to occupy the boundaries and the bulk [see Fig.~\ref{f4}(b)]. Due to the overlap of the evanescent modes between nearest-neighbour waveguides, the photon distribution will evolute to all corners. A 1/4 fractional probability of photon on the four corners can be observed even when we inject a photon in a single corner, which is consistent to the topological corner charge $Q^{(4)}_c=\frac{1}{4}$. It is same for the case of $C_2$ symmetric lattice (see Methods).
	
	In our experiment, we fabricate the lattice with parameters as $d_a^x=d_a^y$ varying from 13 to $13.4\ \mu$m with step of $0.1\ \mu$m, and $d_b^x=d_b^y$ fixed as $11\ \mu$m. We inject the single photon into lattices from one of the lattice corners, the single-photon is not localized in the excited corner as previous results shown in Fig.~\ref{f2} and the probability distribution occupies all four corners and the sites in the lattice boundaries. We confirm and demonstrate the result with four kind lattices with different parameter of $t_a/t_b$ varying from 0.65 to 0.70 and evolution distances as shown in Fig.~\ref{f4}(c). If we investigate the four-corner output probability distribution as the sub-space of the result, we can find that the finite gap effect can provide a way to obtain single-photon superposition state in a lattice.
	
	\paragraph*{Robustness.} The observed corner states are 2D analogy of Jackiw-Rebbi solitons and robust against disorders and perturbations as long as the rotation symmetry and time-reversal symmetry are preserved. From the perspective of topological band theory, for the $C_2$ lattice, the corner states are mid-gap states and separated from the bulk states. Therefore, as long as the disorders does not close the band gaps and breaking the symmetries, the corner states remain unchanged. For $C_4$ symmetric case, the corner states are not in the band gap, however, they are decoupled from the bulk bands and can be excited individually by photons injection on four rods simultaneously.
	
	Meanwhile, to demonstrate the performance of topological corner modes, we fabricate six samples and each sample also is realized in different evolution distance varying from 10 to 30 mm in step of 5 mm. It means that thirty lattices are fabricated and measured in our experiment. Though all the parameters of the system and the fabrication environment during femtosecond laser direct-write process have been locked and optimized, it is still inevitable that the disorder induced by fabrication exists in each lattice. Nevertheless, the performance of corner states in the experimental result almost have no change, implying the distinguishing robust feature of topological phase. 
	
\begin{figure}[htbp]
		\centering
		\includegraphics[width=1.0\linewidth]{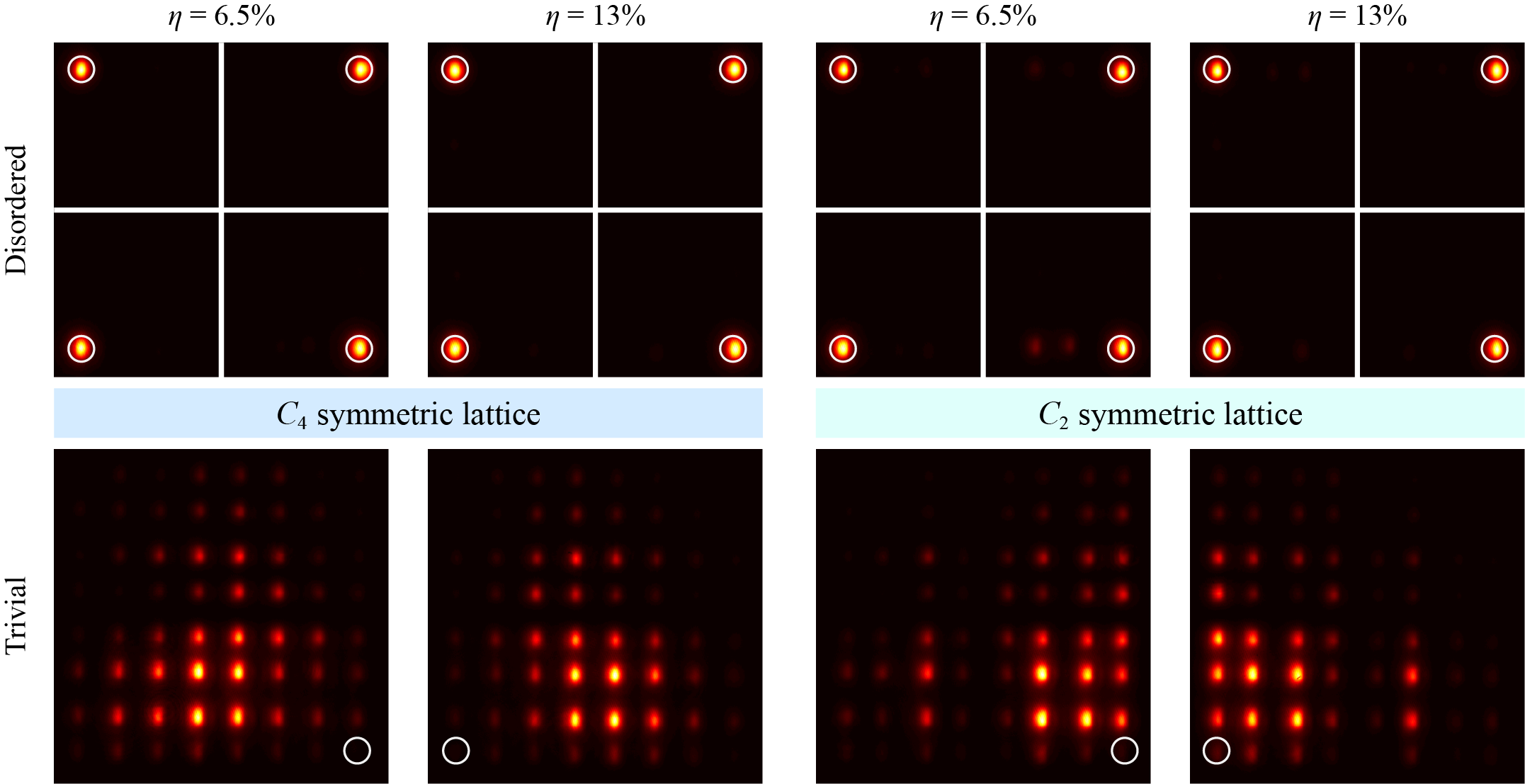}
		\caption{\textbf{The robustness of the photonic lattice.} The photon is still well confined in the excited sites even when we have introduced the disorder in to the lattices, implying the robustness of the explored second-order topological phase in both $C_2$ and $C_4$ symmetric lattices. For the topological trivial cases, the photons can not be confined in the corners and diffuse into the whole lattice. The white circle points out the excited site in the disorder cases.}
		\label{s_robustness}
		
\end{figure}

	Besides, we also introduce the disorder into the pure lattice and demonstrate the robustness of the topological phase. In our experiment, we set the $\eta=\varDelta d/\bar{d}$ as disorder level, where $\varDelta d$ is the introduced disorder separation distance and $\bar{d}$ is the averaged separation distance of the pure lattice. As shown in Fig.~\ref{s_robustness}, the photon is still well confined in the excited sites, implying the robustness of the explored second-order topological phase. We set $d_a^x = d_a^y = 14$ mm, $d_b^x  = d_b^yc  = 18$ mm for $C_4$ symmetric trivial lattice and $d_a^y = d_b^y = 16$ mm, $d_a^x = 14$ mm, $d_b^x  = 18$ mm for $C_2$ symmetric trivial lattice. The photons can not be confined in the corners and diffuse into the whole lattice, as shown in Fig.~\ref{s_robustness}.\\
	
	\subsection*{Conclusion and discussion}
	In summary, we present direct observation of higher-order topological phases and the topological protection to quantum superposition and entanglement in two-dimensional photonic lattice fabricated with femtosecond laser direct writing technique. We demonstrate the different properties of second-order corner modes in the photonic lattice with $C_4$ and $C_2$ symmetry.
	
	We employ single-photon superposition state to identify the corner states in the meanwhile show a topological protection mechanism isolating the quantum entanglement from diffusion-induced decoherence, which provides a promising and valuable resource for quantum information processing.
	
	Our implementation of SOTI at single-photon level can facilitate the way for studying lower dimensional topological localized states with controllable photon numbers in the quantum regime. Unlike recent works on $C_6$~\cite{HOTI10} and $C_3$~\cite{C3} symmetric waveguide arrays, we exhaust the predictions of photonic SOTI in the remaining $C_4$ and $C_2$ symmetric systems, and point out that the corner state can be embedded in the bulk states while being decoupled and can be excited individually from them. Our results extend the protection mechanism of topological phases into quantum regime, and demonstrate the compatibility of quantum photonics and topological phase. The demonstrated key elements, including integrated structures, higher-order topological crystalline insulator, built-in superposition state generation, and heralded single-photon imaging, can enrich the field of quantum topological photonics.\\
	
	\paragraph*{Acknowledgments}
	\noindent The authors thank Roberto Osellame and Jian-Wei Pan for helpful discussions. This research is supported by National Key R\&D Program of China (2017YFA0303700, 2017YFA0303702,2017YFA0304203, 2018YFA0306200, and 2019YFA0308700), National Natural Science Foundation of China (11690033, 61734005, 11761141014, 11604392, 11625418, 11890700, and 51732006), Science and Technology Commission of Shanghai Municipality (17JC1400403), Shanghai Municipal Education Commission (2017-01-07-00-02-E00049), IRT\_17R70, 1331KSC and 111 Project (D18001), China Postdoctoral Science Foundation Funded Project (2019M661784). X.-M.J. acknowledges additional support from a Shanghai talent program.\\
	
	\subsection*{Methods}
	
	\paragraph*{Topological classification and bulk polarization:}
	The time-reversal symmetry of waveguide lattice leads to a vanishing Berry curvature and a zero-Chern number~\cite{SSH2,SSH3}. However, the extra $C_2$ and $C_4$ rotation symmetries will put topologically non-trivial constrains on the eigenfunctions, forming the topological crystalline insulators (TCIs)~\cite{TCI1}. In our cases, the TCIs can be classified by the 2D bulk polarization~\cite{POL1,POL2} defined as follows, 
	\begin{equation}
		P_i=-\frac{1}{(2\pi)^2}\int_{BZ} d^2\bm{k}\mathrm{Tr} [\hat{{\cal A}}_i] , \quad i=x,y
	\end{equation}
	where $BZ$ presents the first Brillouin zone, $(\hat{{\cal A}}_i)_{mn}(\textbf{k})=\mathrm{i}\bra{u_m(\textbf{k})}\partial_{k_i}\ket{u_n(\textbf{k})}$, with $m$, $n$ run over all bands below the gap, $\ket{u_m(\textbf{k})}$ is the periodic part of the eigenfunction for the $m$th band. The 2D polarization is simply related to the 2D Zak phase via $\theta_i=2\pi P_i$ for $i=x,y$.
	
	Besides, the value of the 2D bulk polarization is equal to position of the Wannier center~\cite{WC}. Due to the two mirror symmetries in the waveguide array, the bulk polarization is quantized and the Wannier center is restricted at the maximal Wycoff position of the unit cell. The Wannier center can be applied to investigate the topological classes. If the Wannier center is restricted at the center of the unit cell, namely $(P_x, P_y)=(0, 0)$, the system is adiabatically connected to the atomic insulators which is topologically trivial insulators. Nevertheless, if the Wannier center is restricted to the center of the edge of the unit cell, namely $(P_x, P_y)=(0, \frac{1}{2})$ or $(P_x, P_y)=(\frac{1}{2}, 0)$, it corresponds to the first-order topological insulator with 1D edge states. When $(P_x, P_y)=(\frac{1}{2}, \frac{1}{2})$ which means that the Wannier center is located at the corner of the unit cell, the system is a second-order topological insulator with both 1D edge states and 0D corner states. The topological edge states and corner states are the Jackiw-Rebbi solitons for the two topologically distinct bulks and edges respectively~\cite{soliton}.
	
	For $C_4$ symmetric array, we have $P_x=P_y$ and since $P_x$ can be either $\frac{1}{2}$ or $0$, the bulk polarization forms a $\mathbb{Z}_2$ topological index of the system. However, this is not case for $C_2$ symmetric array because $P_x$ and $P_y$ can be independent to each other. Therefore the bulk polarization forms a $\mathbb{Z}_2\times\mathbb{Z}_2$ topological index for $C_2$ symmetric system. For our cases, when $t^i_a<t^i_b$ for $i=x,y$, we have $(P_x, P_y)=(\frac{1}{2}, \frac{1}{2})$ which implies that the waveguide array is in the SOTI phase.

	\paragraph*{Determining the values of topological indices:} To further determine the values of topological indices, we need to investigate the coupling configurations of our lattice. If we consider two $n-$fold rotation symmetric topological crystalline insulator, the sum of them is also a TCI with the symmetry being the sum of previous $n-$ fold rotation symmetry. This property ensures a free Abelian additive structure of the classification of TCIs and therefore we can choose a set of primitive systems to generate all TCIs up to stable equivalence~\cite{TCI1}. We define these primitive systems as the primitive generators which satisfy certain rotation symmetry. 
	
	For our $C_4$ symmetric lattice, the primitive generator is $h^{(4)}_{1b}$ according to the algebraic method~\cite{TCI1, TCI2}. The previous 2D bulk polarization can be directly obtained from the topological indices as
	\begin{equation}
		P^{(4)}_x=P^{(4)}_y=\frac{1}{2}[X_1],
	\end{equation}  
	and it is defined modulo 1. The corner states arise due to the filling anomaly: the mismatch between the $C_4$-symmetry and conservation of the number of photonic eigenstates and we can define a topological corner index as follows 
	\begin{equation}
		Q^{(4)}_c=\frac{1}{4}([X_1]+2[M^{(4)}_1]+3[M^{(4)}_2]),
	\end{equation}
	and it is also defined modulo 1. In $C_4$ symmetric lattice, for the non-trivial case, we have $[X_1]=-1$, $[M_1]=1$ and $[M_2]=0$. Therefore the bulk polarization is $P^{(4)}_x=P^{(4)}_y=\frac{1}{2}$ which is consist with previous calculations and the topological corner index is $Q^{(4)}_c=\frac{1}{4}$, indicating $\frac{1}{4}$ fractionalized corner states at each of four corners.
	
	For $C_2$ symmetric lattice, the primitive generator of our waveguide array is also $h^{(4)}_{1b}$. The corresponding 2D bulk polarization is now calculated by
	\begin{equation}
		P^{(2)}_x=-\frac{1}{2}([Y_1]+[M^{(2)}_1]),
		P^{(2)}_y=-\frac{1}{2}([X_1]+[M^{(2)}_1]),
	\end{equation} 
	and the topological corner index is defined as
	\begin{equation}
		Q^{(2)}_c=\frac{1}{4}(-[X_1]-[Y_1]+[M^{(2)}_1]).
	\end{equation}
	In $C_2$ symmetric lattice, for the non-trivial case, we have $[X_1]=-1$, $[Y_1]=-1$ and $[M^{(2)}_1]=0$. Therefore the bulk polarization is $P^{(2)}_x=P^{(2)}_y=\frac{1}{2}$ which is consist with previous calculations. However, in this case, the topological corner index is $Q^{(2)}_c=\frac{1}{2}$, indicating $\frac{1}{2}$ fractionalized corner states at each of four corners which is different from the $C_4$ symmetric lattice.
	
	\paragraph*{Fabrication and measurement of the lattices on a photonic chip:} We fabricate the samples in alkaline earth boro-aluminosilicate glass substrate (refractive index $n_0=1.514$ for the writing laser at a wavelength of 513 nm) using the laser system operating at a repetition rate of 1 MHz and a pulse duration of 290 fs. The light is focused inside the sample with a 50X microscope objective (NA=0.50) after being reshaped with a spatial light modulator. We continuously move the substrates using a high-precision three-axis translation stage with a constant velocity of 10 mm/s to create the lattices by the laser-induced refractive index increase.
	
	In the experiment, we inject the photons into the input waveguides in the photonic chip using a 20X objective lens. After a total propagation distance through the lattice structures, the outgoing photons are first collimated with a 10X microscope objective, then detected and analyzed by a combination of wave plates and polarizers.
	
	\paragraph*{The generation and imaging of the heralded single-photon state:} The single-photon source with the wavelength of 810 nm are generated from periodically-poled KTP (PPKTP) crystal via type-II spontaneous parametric down conversion. The generated photon pairs are separated to two components, horizontal and vertical polarization, after a long-pass filter and a polarized beam splitter (PBS). One should notice that the measured patterns would come from the thermal-state light rather than single photons if we inject only one polarized photon into the lattices without external trigger. Therefore, we inject the horizontally polarized photon into the lattices, while the vertically polarized photon acts as the trigger for heralding the horizontally polarized photons out from the lattices with a time slot of 10 ns. The measured second-order anti-correlation parameter is 0.026$\pm$0.003, implying that a single photon is well preserved in the corner states. We capture each evolution result using the ICCD camera after accumulating in the external mode for 600s.
	
	\paragraph*{The 3D 1$\times$4 photonic coupler:}
	The Hamiltonian of the 3D 1$\times$4 photonic coupler could be written as
	\begin{equation}         
		H=
		\left(                 
		\begin{array}{ccccc}   
			0 & c & c & c & c\\  
			c & 0 & 0 & 0 & 0\\  
			c & 0 & 0 & 0 & 0\\  
			c & 0 & 0 & 0 & 0\\  
			c & 0 & 0 & 0 & 0\\ 
		\end{array}
		\right),
		\label{Hc}
	\end{equation}
	where the entry waveguide is labeled as 1 and the other four waveguides are labeled as 2 to 5 respectively, $c$ is the coupling strength. In our experiment, the distance of 1$\times$4 photonic coupler is set as $L = \frac{\pi}{4c}$. According to the evolution operator $U=e^{-iHL}$, we can obtain
	\begin{equation}         
		U=
		-\frac{1}{4}\left(                 
		\begin{array}{ccccc}   
			0 & 2i & 2i & 2i & 2i\\  
			2i & -3 & 1 & 1 & 1\\  
			2i & 1 & -3 & 1 & 1\\  
			2i & 1 & 1 & -3 & 1\\ 
			2i & 1 & 1 & 1 & -3\\  
		\end{array}
		\right).
		\label{coupler}
	\end{equation}
	When we inject the single photon in to the entry waveguide, then $\ket{\psi_{in}}=[1\ 0\ 0\ 0\ 0]^T$. According to $\ket{\psi_{out}}=U\ket{\psi_{in}}$, we obtain $\ket{\psi_{out}}=\frac{i}{2}[0\ 1\ 1\ 1\ 1]^T$, implying that the probability and phase of single photon in four waveguides are uniform.
	
	\paragraph*{The band gap of $C_2$ symmetric lattice:}
	For $|t^i_a/t^i_b| \ll 1$ in $C_2$ symmetric lattices, the band gap is large and the corner states are well separated from the bulk states. The real space field distributions of corner states decay away from the corner position in forms of $e^{-m\delta x}$ and $e^{-m\delta y}$ along $x$ and $y$ directions where $m$ stand for the mass terms in the effective Hamiltonian which is proportional to the size of the band gaps. $\delta x$ and $\delta y$ represent the distances from the corner positions along $x$ and $y$ directions respectively. Therefore, in this case, the photons are well confined at four corners. On contrary, when $|t^i_a/t^i_b|$ is gradually increased to 1, the size of band gaps reduced and the localization of corner states are weakened, leading to field distributions in the edge and bulk of the lattices.

	\clearpage
	
	\clearpage
	
	\clearpage
	
	\clearpage
	
	\clearpage
	
	\clearpage

\newpage

\section*{Supplementary Materials: Protecting Quantum Superposition and Entanglement with Photonic Higher-Order Topological Crystalline Insulator}
\baselineskip18pt
\noindent Yao Wang,$^{1,2,\dagger}$ Bi-Ye Xie,$^{3,4,\dagger}$ Yong-Heng Lu,$^{1,2}$ Yi-Jun Chang,$^{1,2}$\\
Hong-Fei Wang,$^{3,4}$ Jun Gao,$^{1,2}$ Zhi-Qiang Jiao,$^{1,2}$ Zhen Feng,$^{1,2}$ Xiao-Yun Xu,$^{1,2}$\\
Feng Mei,$^{5,6}$ Suotang Jia,$^{5,6}$ Ming-Hui Lu,$^{3,4,7,8}$ and Xian-Min Jin$^{1,2}$\\
\\
$^1$Center for Integrated Quantum Information Technologies (IQIT), School of Physics and Astronomy and State Key Laboratory of Advanced Optical Communication Systems and Networks, Shanghai Jiao Tong University, Shanghai 200240, China\\
$^2$CAS Center for Excellence and Synergetic Innovation Center in Quantum Information and Quantum Physics, University of Science and Technology of China, Hefei, Anhui 230026, China\\
$^3$National Laboratory of Solid State Microstructures, Nanjing University, Nanjing 210093, China\\
$^4$Department of Materials Science and Engineering, Nanjing University, Nanjing 210093, China\\
$^5$State Key Laboratory of Quantum Optics and Quantum Optics Devices, Institute of Laser Spectroscopy, Shanxi University, Taiyuan, Shanxi 030006, China\\
$^6$Collaborative Innovation Center of Extreme Optics, Shanxi University, Taiyuan, Shanxi 030006, China\\
$^7$Jiangsu Key Laboratory of Artificial Functional Materials, Nanjing 210093, China\\
$^8$Collaborative Innovation Center of Advanced Microstructures, Nanjing University, Nanjing 210093, China\\
$^\dagger$These authors contributed equally to this work\\

\baselineskip24pt

\subsection*{The quantum evolution in the waveguide array}\label{quantum_evolution}
In this section, we will discuss the quantum evolution of photons in the waveguide array, giving the description on the features of corner states in our work. In our system, the dynamic behavior of photon is governed by evolution equation, obtained from paraxial wave equation by employing the tight-binding approximation, as
\begin{align}
i\partial_z\psi_n&=-t\left(\psi_{n-1}+\psi_{n+1}\right)-\beta \psi_{n}\label{H1}\\
&=H\psi_{n}\label{H2},
\end{align}
where the $t$ is the coupling coefficient between the adjacent sites and $\beta$ is the on-site energy. According to the quantum mechanics, the evolution of photon in the system obeys the equation as
\begin{equation}
\psi(t)=e^{-iHt}\psi(0),
\label{evolution}
\end{equation}
where $\psi(0)$ is the initial wavefunction of photon, and $\psi(t)$ is the wavefunction after evolution time $t$. We decompose the initial wavefunction in components of all eigen states as $\psi(0) =\sum_j c_j \ket{\phi}_{j}$, where $c_j$ is the probability amplitude of eigen state $\ket{\phi}_{j}$. According to Eq.~\ref{evolution}, we can find that
\begin{align}
\psi(t)&=e^{-iHt} \sum_j c_j \ket{\phi}_j\\
&= \sum_j c_j e^{-iE_{j}t}\ket{\phi}_j,
\end{align}
where $E_j$ is the eigen energy of the eigen state $\ket{\phi}_{j}$. Now, we can get the probability amplitude proportion of $\ket{\phi}_{j}$ and $\ket{\phi}_{k}$ as 
\begin{equation}
\eta(t) = c_j/c_ke^{-i(E_j-E_k)t}.
\label{probability}
\end{equation}
It is obvious that the probability amplitude proportion is maintained if $E_j=E_k$. Meanwhile, if $E_j\neq E_k$, due to $|\eta(t)|^2=|\eta(0)|^2=|c_j/c_k|^2$, the probability amplitude proportion is still maintained though there is the relative phase between them if $c_j$ and $c_k$ are the real numbers.

In the following, we analyze the property of the probability amplitude $c_j$ in our work. As described in Eq.~\ref{H1}, the $H$ in the Eq.~\ref{H2} satisfies $H^{\dagger}=H$, implying that the eigen-values $E_j$ and eigen-states $\ket{\phi}_j$ of $H$ are real. For the case of exciting the lattice from only one site, at $t=0$, $\psi(0)=c_j\ket{\phi}_j$, where the $c_j$ is real number. For the case of exciting the lattice with single-photon superposition state, the initial state is also real due that the four modes obtained by the 3D 1$\times$4 photonic coupler are in the same phase. Such that, the probability amplitude $c_j$ is also the real number in our work, implying that the probability amplitude proportion of the corner states is maintained with the evolution time.

\begin{figure}[htbp]
	\centering
	\includegraphics[width=0.7\linewidth]{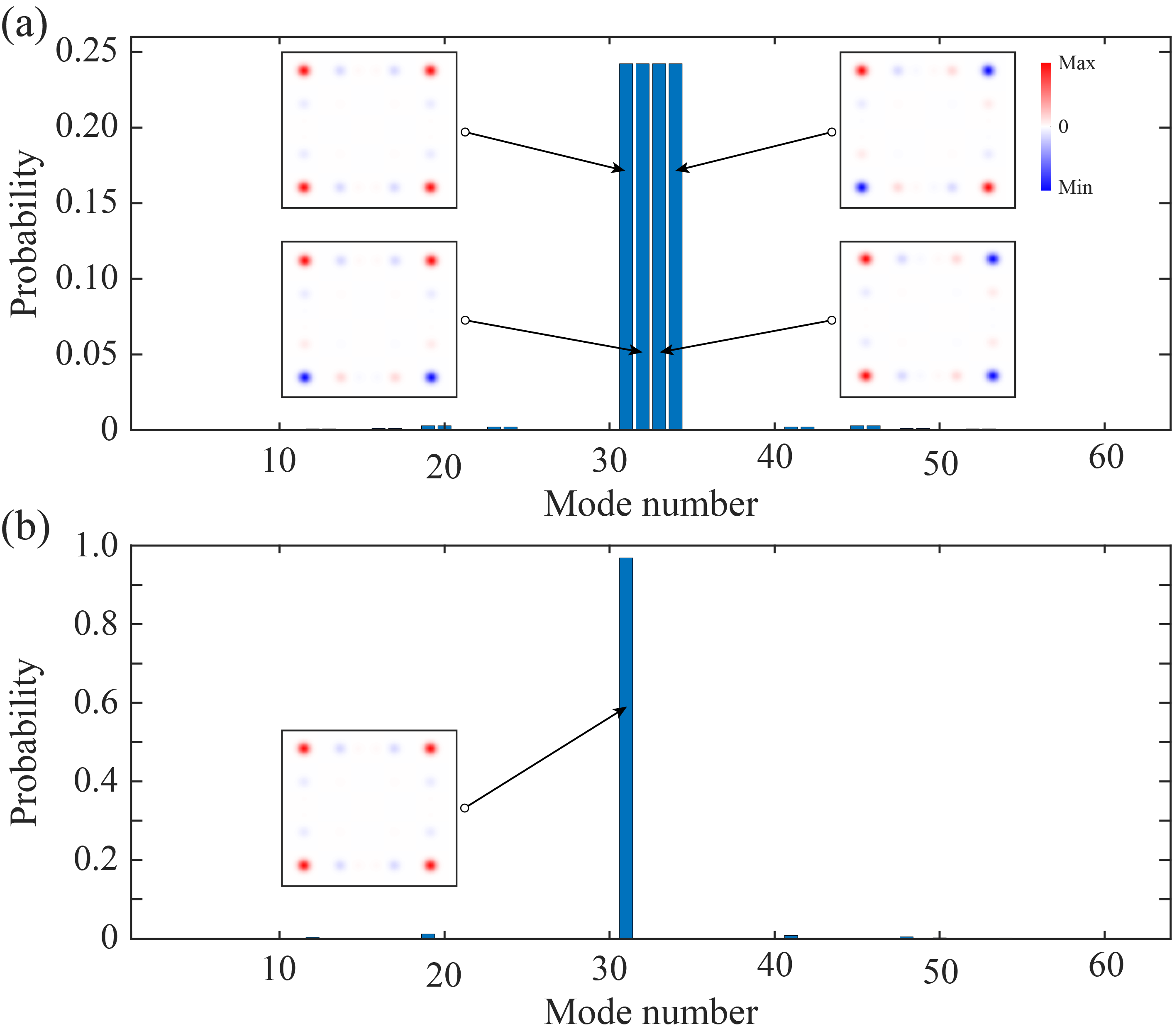}
	\caption{\textbf{Probability amplitude proportion of the excited corner states in $C_2$ lattice.} \textbf{(a)} Proportion of excited modes when exciting the lattice from one corner. \textbf{(b)} Proportion of excited modes when exciting the lattice with the single-photon superposition state.}
	\label{s_identify}
\end{figure}

In our experiment, for the $C_2$ symmetric lattice, only four corner states are excited simultaneously when we excite the lattice from one site in the corner, as shown in Fig.~\ref{s_identify}(a). The energies of the corner states are nearly degenerate, according to Eq.~\ref{probability}, the probability amplitude proportion and the relative phase of the corner states are maintained. Though the relative phase between the corner states and other trivial states would change with the evolution time, the single-photon distribution will still confine on the corner due to the high probability amplitude proportion of eigen corner states. In the other word, the probability distribution of single photon could be maintained in the excited site. This property is different from the all-dielectric photonic crystals, in which the field distribution would occupy the four corners even that the lattice is excited from just one corner. Meanwhile, only the corner mode with zero relative phase among the corners is excited when we excite the lattice by the single-photon superposition state, as shown in Fig.~\ref{s_identify}(b).

\begin{figure}[htbp]
	\centering
	\includegraphics[width=0.7\linewidth]{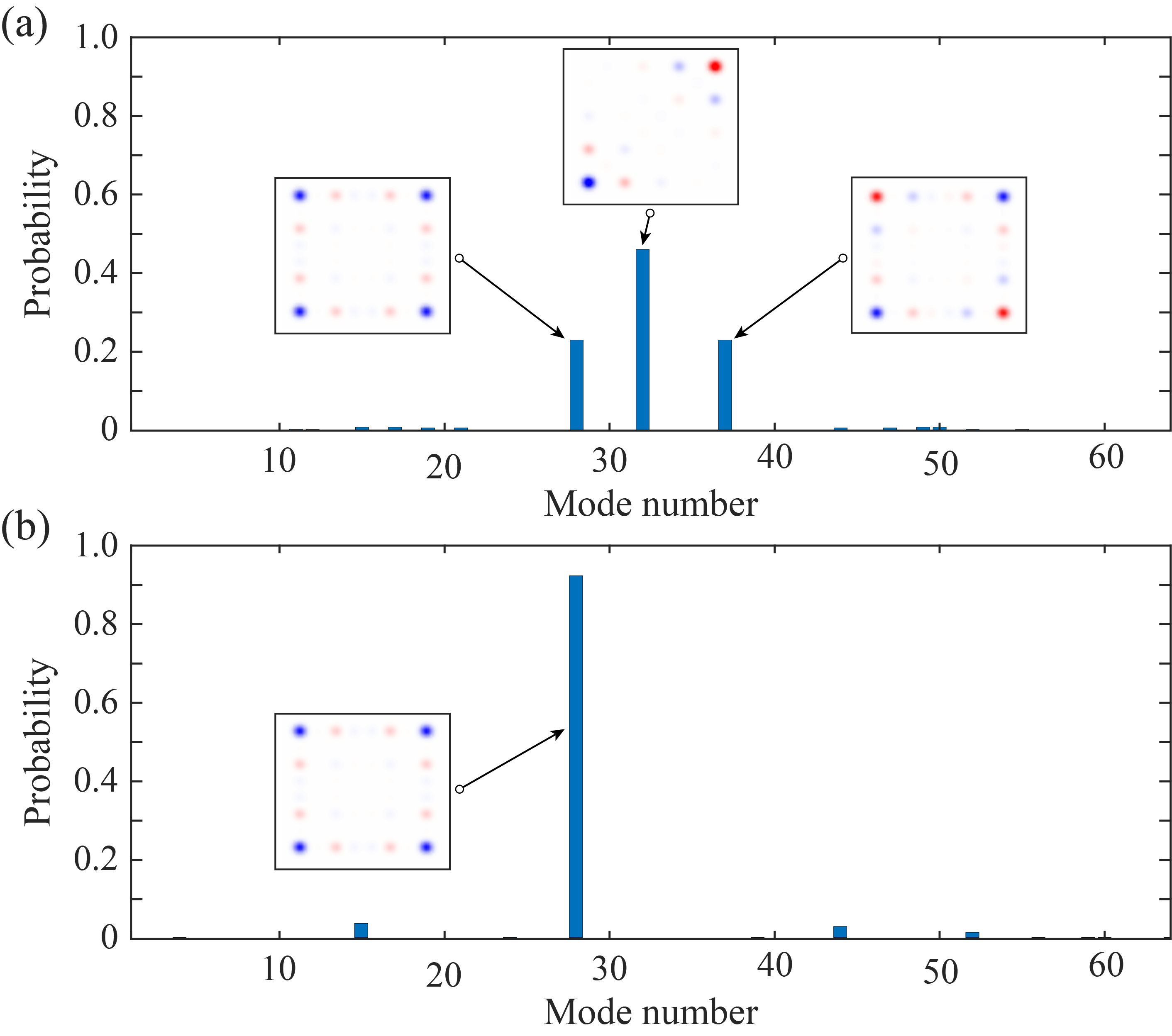}
	\caption{\textbf{Probability amplitude proportion of the excited corner states in $C_4$ lattice.} \textbf{(a)} Proportion of excited modes when exciting the lattice from one corner. \textbf{(b)} Proportion of excited modes when exciting the lattice with the single-photon superposition state.}
	\label{s_identify_c4}
\end{figure}

For the $C_4$ symmetric lattice, there are three types nearly degenerate eigen corner states, all of them are excited when we excite the lattice from one corner, as shown in Fig.~\ref{s_identify_c4}(a). Similar to the $C_2$ symmetric lattice, only the corner state with zero relative phase among the corners is excited when we excite the lattice by the single-photon superposition state, as shown in Fig.~\ref{s_identify_c4}(b).

\begin{figure}[htbp]
	\centering
	\includegraphics[width=1.0\linewidth]{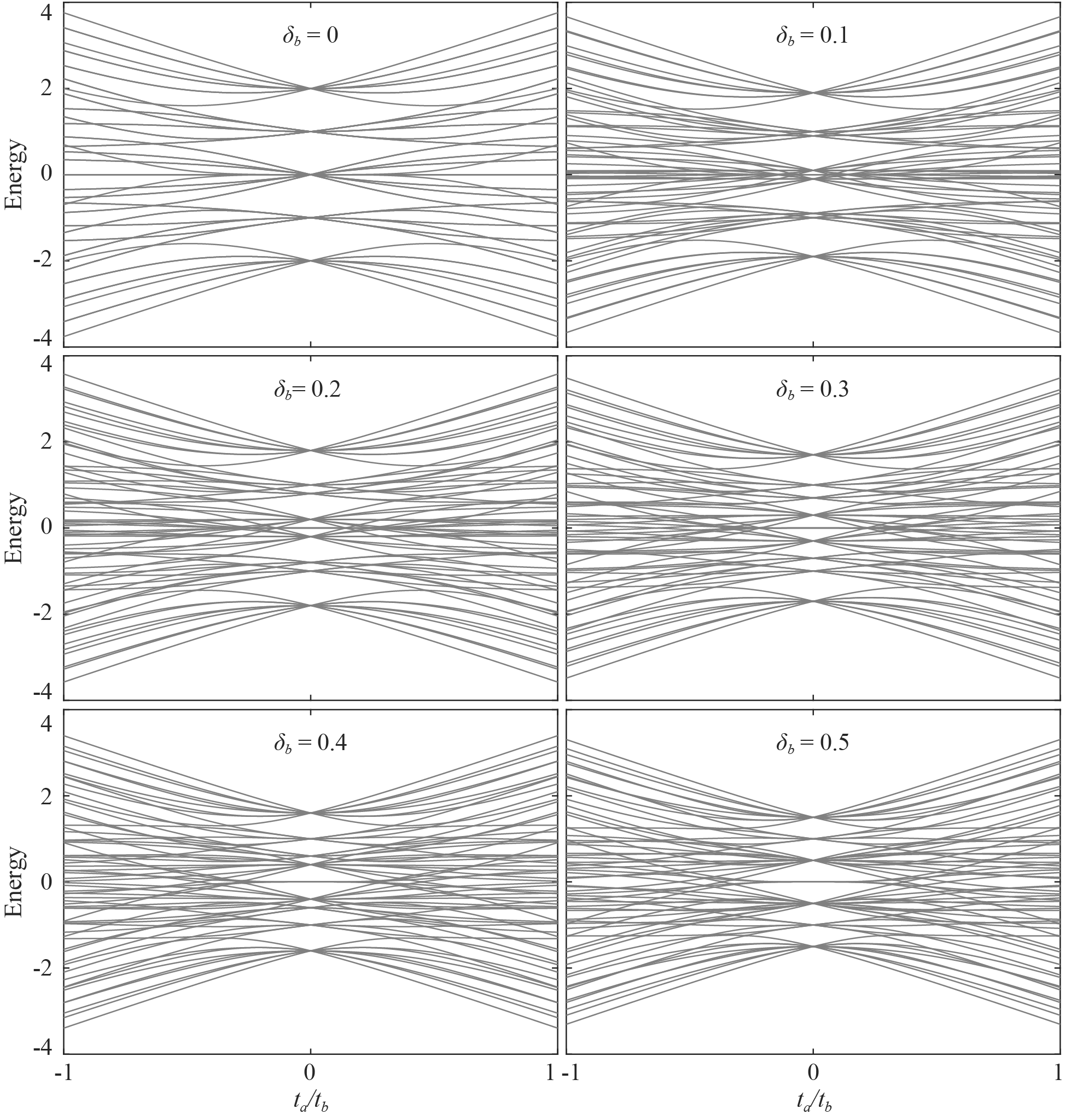}
	\caption{\textbf{Band spectrum of the lattice with different $\delta_b$.}}
	\label{s_band}
\end{figure}

\begin{figure}[htbp]
	\centering
	\includegraphics[width=1.0\linewidth]{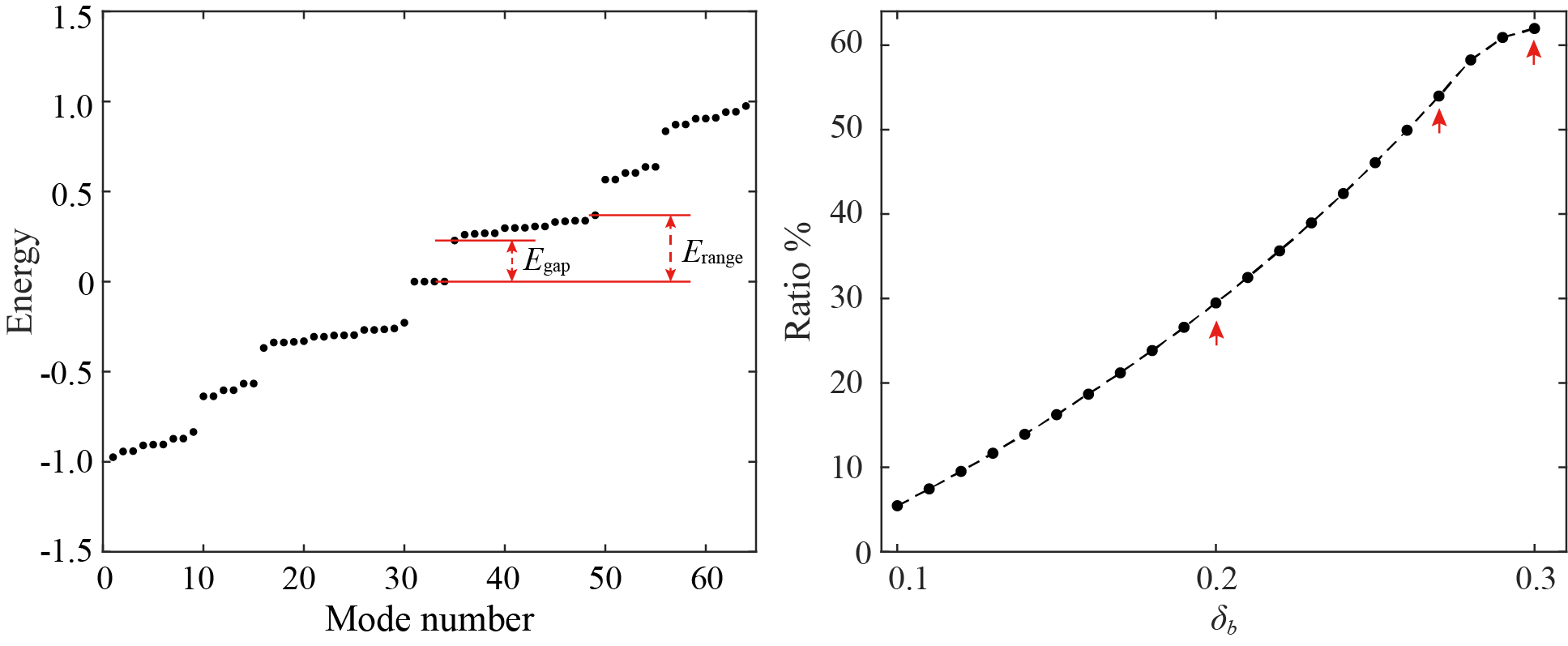}
	\caption{\textbf{The ratio of the band gap size with different $\delta_b$.} The red arrows point out the parameters adopted in our experiment.}
	\label{s_gap}
\end{figure}

\subsection*{Band spectrum and symmetry}	
We define $\delta_a=|t_a^x-t_a^y|$ and $\delta_b=|t_b^x-t_b^y|$ to characterize the symmetry of lattice. The lattice is in $C_4$ symmetry if  $\delta_a=\delta_b=0$, otherwise the lattice is in $C_2$ symmetry. As shown in Fig.~\ref{s_band}, we set $\delta_a=0$ and modulate the $\delta_b$, the degenerate modes divide, which renders the zero-energy corner modes to be gaped by the bands. The gap becomes larger with the increasing of $\delta_b$. We define the ratio of the band gap size as the $E_{gap}/E_{range}$ as shown in Fig.~\ref{s_gap}, which is quite large in photonic crystals.

\begin{figure}[htbp]
	\centering
	\includegraphics[width=1.0\linewidth]{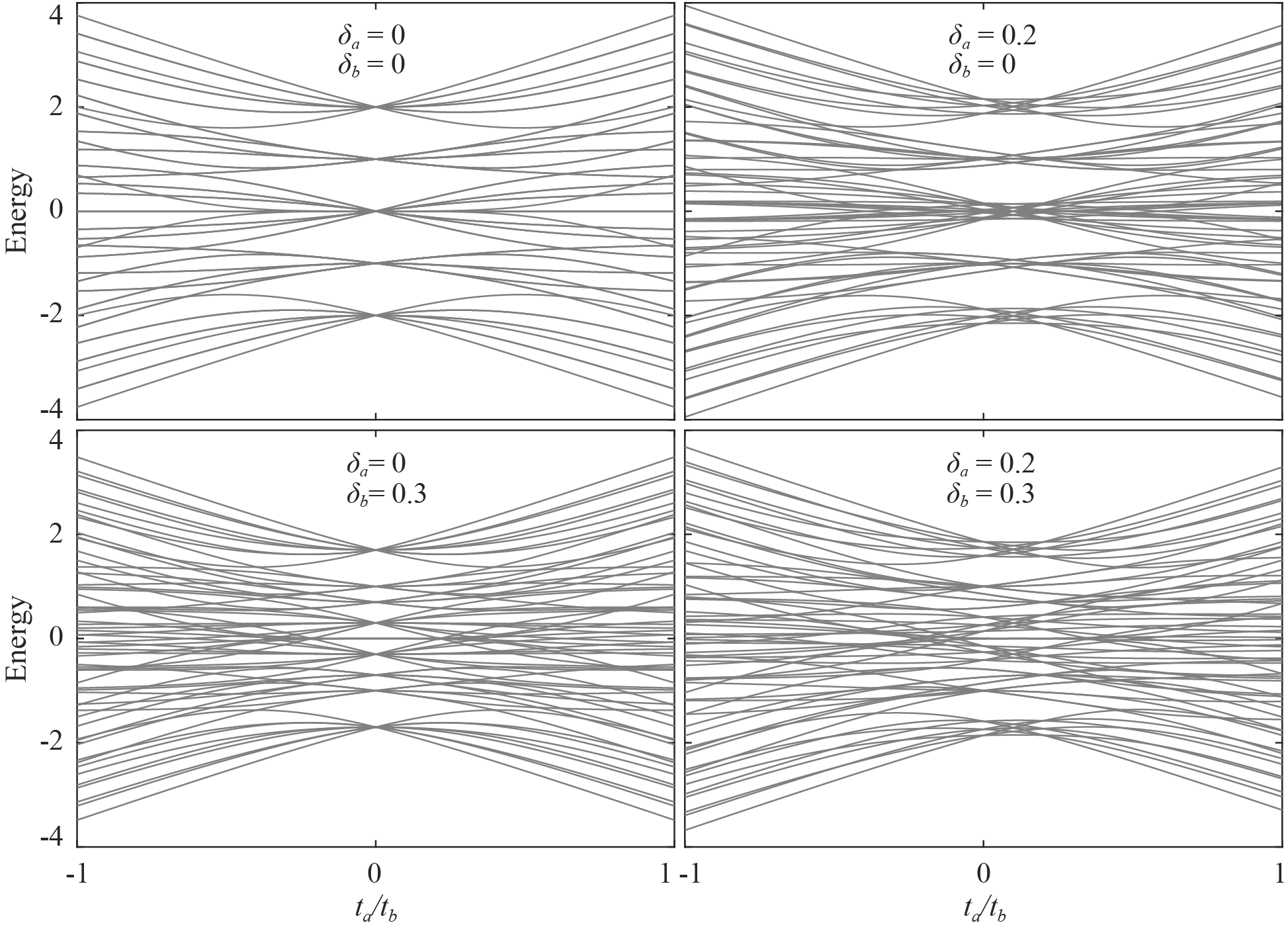}
	\caption{\textbf{Band spectrum of the lattice with different $\delta_a$ and $\delta_b$.}}
	\label{s_band_ab}
\end{figure}

In Fig.~\ref{s_band_ab}, we further compare the band spectrum with different $\delta_a$ and $\delta_b$, the degenerate modes are further divided while the zero-energy corner modes are still in the gap. It should be noted that the opened gap for the corner modes is very small when we modulate $\delta_a$ and keep $\delta_b=0$. The result implies the strong influence of $\delta_b$ on the band spectrum.

\subsection*{Corner modes in $C_2$ symmetric lattice}	
In the main text, we have shown the finite gap effect taking the $C_4$ symmetric lattice for example. As we stated, the phenomenon is also the same with $C_2$ symmetric lattice. In this section, we will give the behind physical mechanism using the theory of quantum mechanics taking the $C_2$ symmetric lattice for example.

\begin{figure}[htbp]
	\centering
	\includegraphics[width=1.0\linewidth]{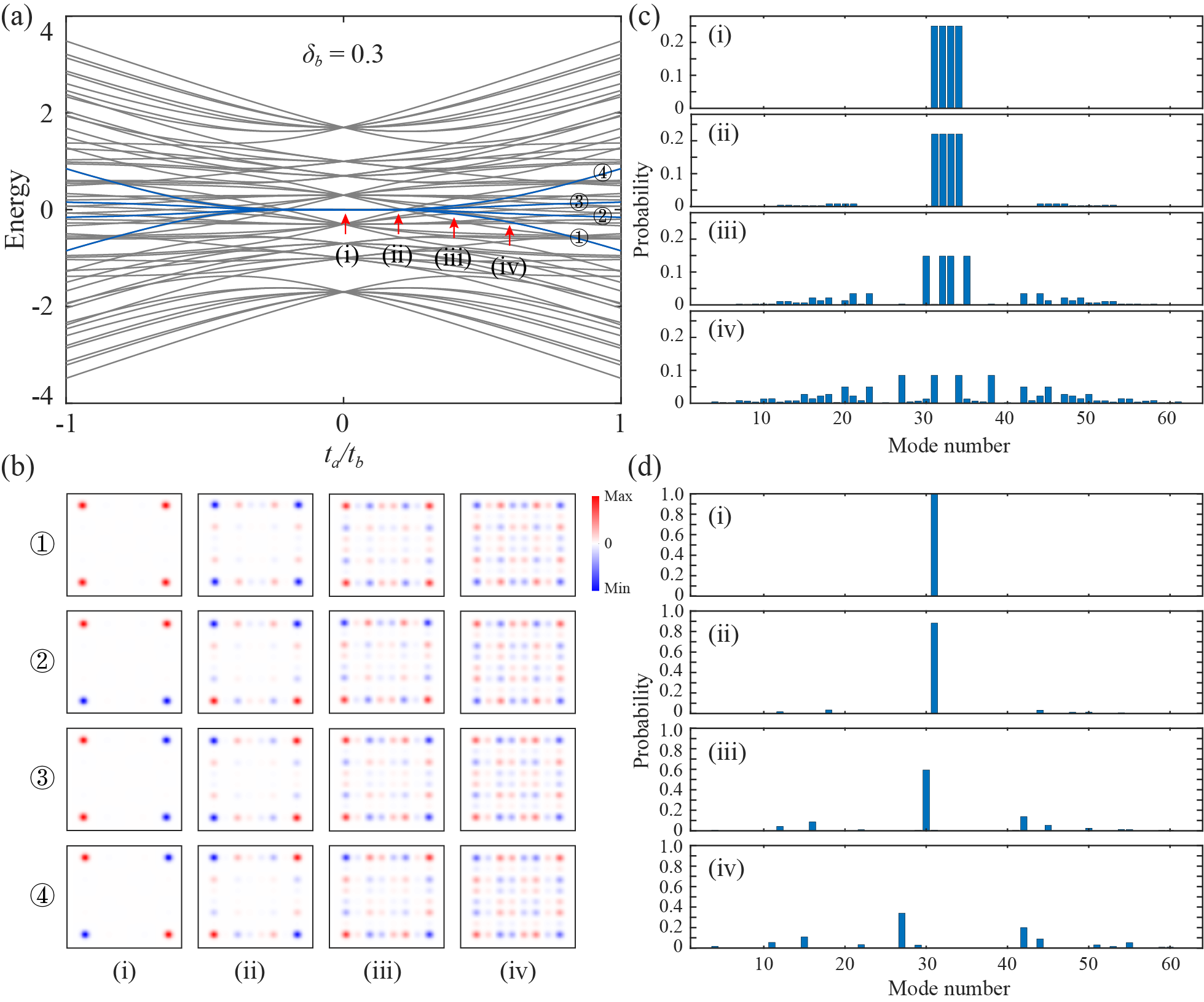}
	\caption{\textbf{Corner modes in $C_2$ symmetric lattice.}}
	\label{s_degenerate}
\end{figure}

As shown in Fig.~\ref{s_degenerate}, the degenerate corner modes transit to non-degenerate with the increase of $t_a/t_b$ for the topological phase [Fig.~\ref{s_degenerate}(a)], and the corresponding spatial distributions become non-localization [Fig.~\ref{s_degenerate}(b)], especially, when the corner modes are not gaped. We also can find that the probability of corner modes is maintained when the them are gaped. The probability decreases for the larger $t_a/t_b$ while trivial modes are also excited, as the result shown for the case of exciting the lattice from one corner [Fig.~\ref{s_degenerate}(c)] and the case of exciting the lattice using the single-photon superposition state [Fig.~\ref{s_degenerate}(d)].

\begin{figure}[htbp]
	\centering
	\includegraphics[width=0.7\linewidth]{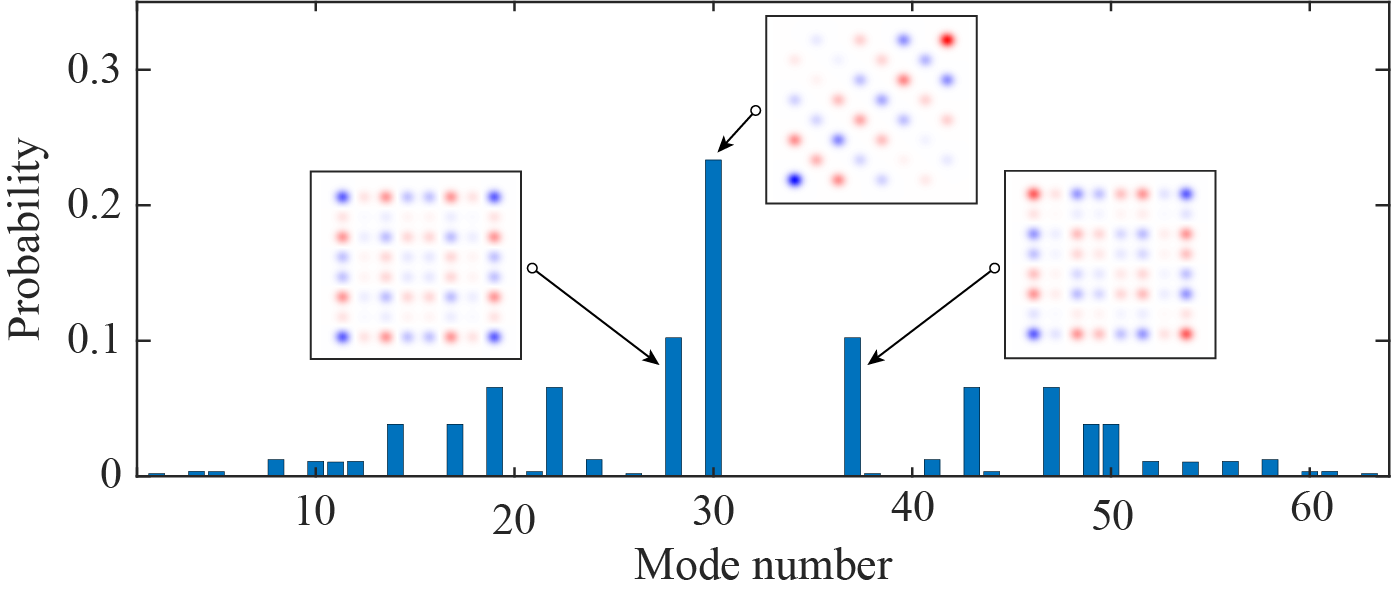}
	\caption{\textbf{Proportions of all the excited modes for the $C_4$ symmetric lattice in experiment.}}
	\label{s_lattice_1to4}
\end{figure}

In the main text, the experiment results show that the photon distribution will evolve to all corners even we inject the photon from the one corner, which seems to contradict the results in Section~\ref{quantum_evolution}. In fact, the result is also can be derived from Eq.~\ref{probability}. Due that the corner modes become non-degenerate, the relative phase between the corner modes is not zero with the evolution, which rendering the photon distribution is not constant. Such that, the photon is able to distribute in the other corners. It should be noted that the proportions of all the excited modes are still not change with the evolution. As shown in Fig.~\ref{s_lattice_1to4}, we give the proportions of all the excited modes for the $C_4$ symmetric lattice in the main text. The corner states are dominating, the photon is able to evolve to other corners by the excited trivial modes and the non-degenerate corner modes.

\subsection*{Corner modes in $C_4$ symmetric lattice}
Although the corner states coexist with the continuum of bulk and edge states, they are not hybridized with these degenerate states. This non-hybridization reveals the topological feature and is protected by the $C_4$ symmetry and chiral symmetry of the lattice. In fact, these corner states are actually the higher-order version of bound states in the continuum (BIC), and they can be separated from the bulk states even under small perturbations as long as the chiral symmetry and C4 symmetry is preserved~\cite{BIC1,BIC2}. The topological protection of BIC maintains even when we introduce non-Hermitian on-site energy in all of the lattice sites except for those of the corners. In this case, the energies of the bulk state will have non-vanishing imaginary parts while the energies of the four degenerate corner states are nearly real and approach to a zero-imaginary component exponentially fast with increasing systems size (see Fig. 2 in Ref.~\cite{BIC1}). Moreover, the corner states as BIC are exactly topologically protected by the same symmetries that protect topological crystalline insulators. 

\subsection*{The entanglement in quantum walk}
\begin{figure}[htbp]
	\centering
	\includegraphics[width=0.4\linewidth]{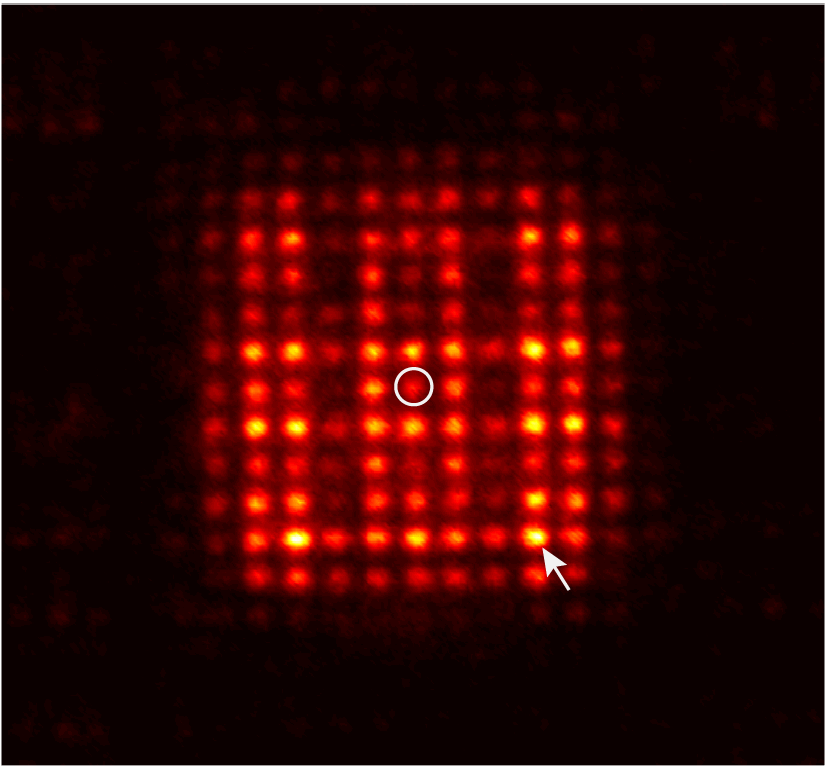}
	\caption{\textbf{The photon distribution of the quantum walk lattice.} The white circle points out the excited site and the white arrow indicates the site we measure the outgoing entangled photon.}
	\label{s_QW}
\end{figure}

In this section, we give the photon distribution of the large-scale two-dimension quantum walk lattice, as shown in Fig.~\ref{s_QW}.

\subsection*{The difference with all-dielectric lattice}
In the all-dielectric lattice, when the frequency approaches zero, the dispersion is linear around the Brillouin zone center, which inevitably breaks the chiral symmetry of the system while this is not the case for the femtosecond-laser direct writing waveguide arrays (FDWWA). The reason behind this phenomenon is that the evanescent waves exponentially decay away from a lattice site in the FDWWA. While they are not for all-dielectric photonic crystals. In all-dielectric photonic crystals, for the lower-frequency band (where the corner states emerge), the eigenmodes are plane-wave like while for higher-frequency bands where the wavelength is comparable to the lattice constant the transmission is realized mainly by the transfer between the resonance modes~\cite{TB}. Therefore, there is a good approximation for the FDWWA to the tight-binding model in all frequency while this is not true for the lower-frequency band in the all-dielectric photonic crystal.

\subsection*{The difference with quadrupolar insulators}
As we have discussed in our manuscript, our implementation of the higher-order topological insulators is very different from the quadrupolar insulators as in Refs.\cite{QI,HOTI3}. In fact, the higher-order topological insulators in our work are a kind of topological crystalline insulators and therefore are topologically distinct for $C_4$ and $C_2$ lattice. The asymmetric coupling here is used to break the $C_4$ symmetry to $C_2$ symmetry and the symmetry group representation at high symmetry points in momentum space is different for $C_4$ and $C_2$ symmetric lattice~\cite{QI2}. Nevertheless, in both cases, the corner states emerge and topologically protected by the filling anomaly. Because of the different topological crystalline phases, the topological corner index and fractional charge between $C_4$ and $C_2$ lattice are different from each other as discussed in the main text and in Ref.~\cite{QI2}. Our work firstly shows the difference between the higher-order topological crystalline insulators in $C_4$ and $C_2$ lattices.

\subsection*{Experimental results}
\begin{figure}[htbp]
	\centering
	\includegraphics[width=0.95\linewidth]{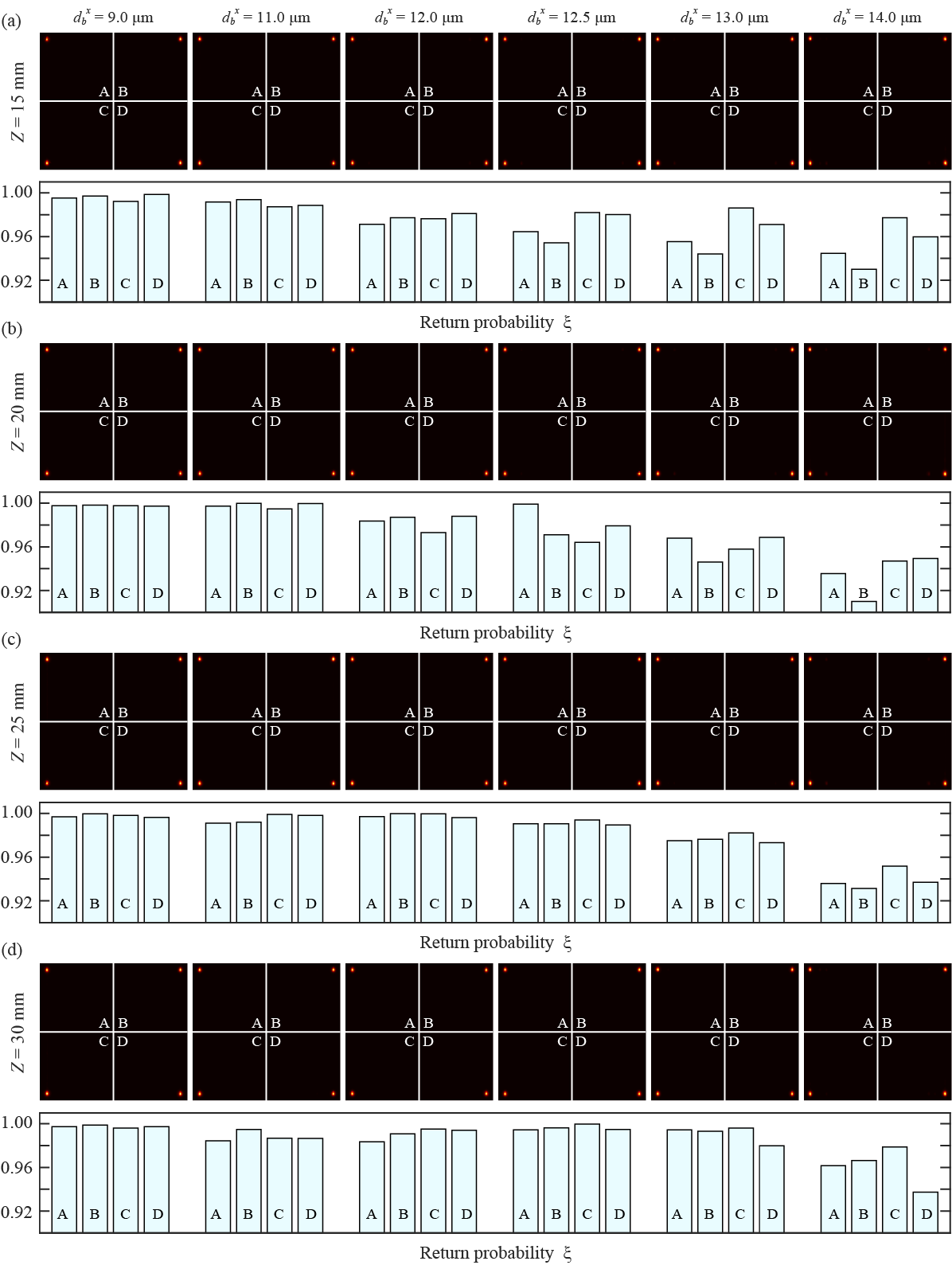}
	\caption{\textbf{Measured single-photon distribution probability and return probability of corner states.}}
	\label{s1}
\end{figure}

\begin{figure}[htbp]
	\centering
	\includegraphics[width=1.0\linewidth]{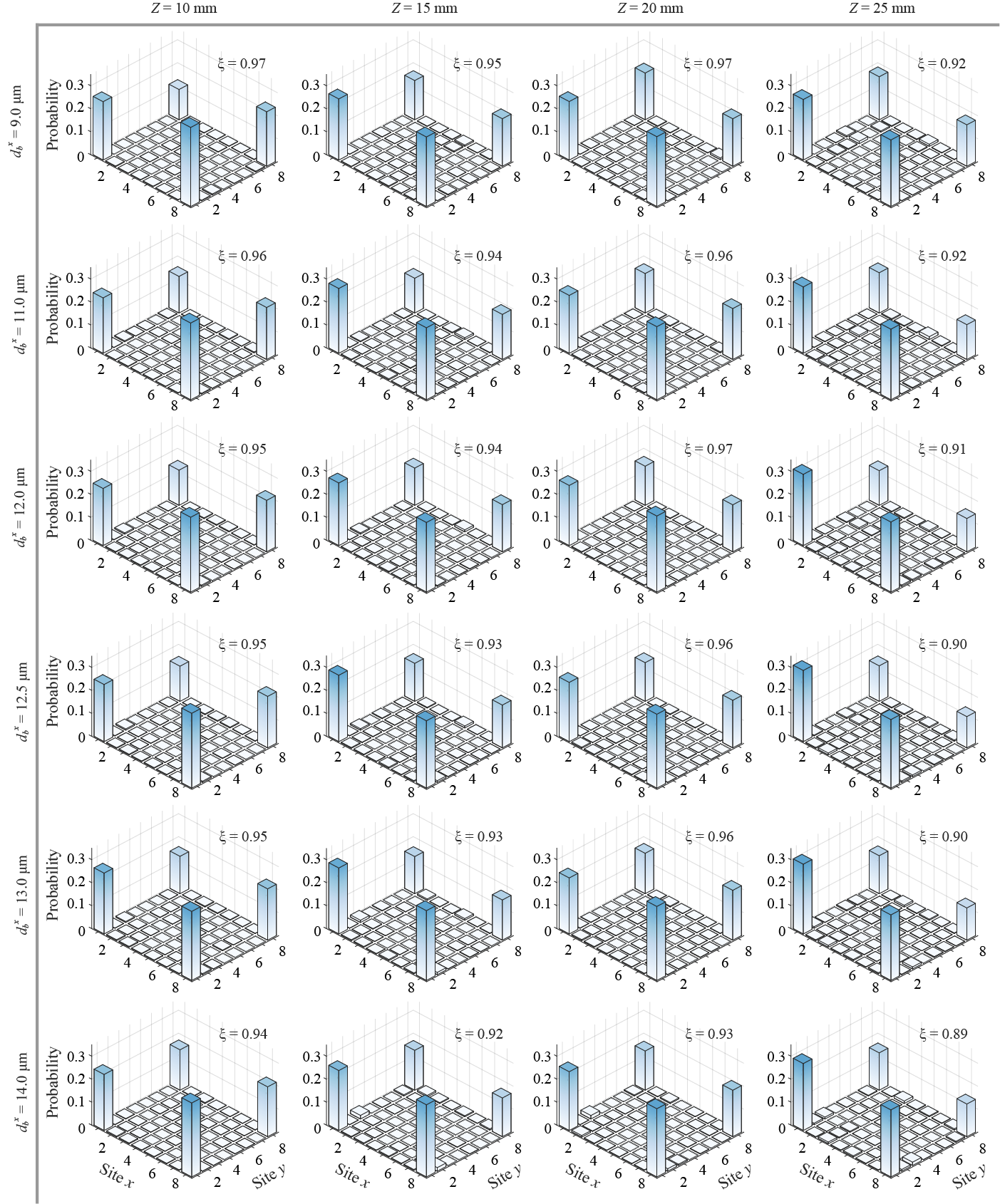}
	\caption{\textbf{Experimental results of corner states by exciting the lattices with single-photon superposition state.}}
	\label{s2}
\end{figure}

We show all the experimental result for all fabricated samples in this section, including the measured single-photon distribution probability and return probability of corner states in Fig.~\ref{s1}, and the experimental results of corner states by exciting the lattices with single-photon superposition state in Fig.~\ref{s2}.

\clearpage

\end{document}